\documentclass[twocolumn,final,twoside,journal]{IEEEtran}
\usepackage{epsfig,amsfonts,color,amsbsy}
\usepackage[nolist]{acronym}
\usepackage{graphicx,cite,amssymb,amsmath,perso}
\usepackage{tikz}
\usepackage{multirow}
\usepackage{bigstrut}
\usepackage{psfrag}
\usetikzlibrary{shapes,arrows}
\usepackage{xcolor}
\usepackage{mathtools}
\mathtoolsset{showonlyrefs}
\usepackage{caption}
\usepackage{subcaption}

\newcommand{\figw}{0.97\columnwidth}
\newcommand{\figwidth}{1\columnwidth}

\newcommand{\todo}[1]{}
\allowdisplaybreaks

\begin{document}
\newcommand{\vecu}{\mathbf{u}}
\newcommand{\vecv}{\mathbf{v}}
\newcommand{\vecx}{\mathbf{x}}
\newcommand{\vecU}{\mathbf{U}}
\newcommand{\vecV}{\mathbf{V}}
\newcommand{\vecX}{\mathbf{X}}

\newcommand{\veca}{\mathbf{a}}

\newcommand{\hw}{w_{\mathsf H}}
\renewcommand{\deg}{\mathrm{deg}}

\newcommand{\field}{\mathbb{F}_2}

\newcommand{\rate}{\mathsf{r}}
\newcommand{\ro}{\rate_{\mathsf o}}
\newcommand{\ri}{\rate_{\mathsf i}}

\newcommand{\ensemble}{\msr{C}}
\newcommand{\oensemble}{\msr{C}_{\mathsf o}}
\newcommand{\code}{\mathbb{C}}

\newcommand{\dmax}{\mathrm{d}_{\mathrm{max}}}
\newcommand{\dmint}{\delta^{\star}}
\newcommand{\dmintt}{d_{\mathrm{min}}^{\star}}
\newcommand{\avgd}{\bar{\Omega}}
\newcommand{\dmin}{d_{\mathrm{min}}}
\newcommand{\G}{\mathsf{G}}
\newcommand{\D}{D}
\newcommand{\Z}{Z}

\newcommand{\we}{A}
\newcommand{\weo}{A^{\mathsf o}}
\newcommand{\wei}{A^{\mathsf i}}
\newcommand{\geo}{G^{\mathsf o}}

\newcommand{\p}{p}
\newcommand{\pl}{\p_{\l}}
\newcommand{\pjl}{ \p_{j,\l} }
\newcommand{\pnl}{ \p_{\nl} }
\newcommand{\npnl}{ \np_{\nl} }
\newcommand{\pjnl}{ \p_{j,\nl} }
\newcommand{\pnlo}{ \p_{\nlo} }
\newcommand{\plo}{ \p_{\lo} }
\newcommand{\np}{\varrho}

\newcommand{\nd}{\delta}
\renewcommand{\d}{d}
\newcommand{\ds}{d^{\star}}
\renewcommand{\l}{l}
\newcommand{\nl}{\lambda}
\newcommand{\nls}{\lambda^\star}
\newcommand{\Hb}{\mathsf H_{\mathsf b}}
\newcommand{\tw}{\tilde w}
\newcommand{\two}{\tilde w_0}
\newcommand{\nlo}{\lambda_0}

\newcommand{\fmax}{\mathsf f_{\textrm{\textnormal{max}}}}
\newcommand{\f}{\mathsf f}
\newcommand{\gmax}{\mathsf g_{\textrm{\textnormal{max}}}}
\newcommand{\g}{\mathsf g}

\newcommand{\lo}{\l_0}

\newcommand{\inner}{\msr{I}}
\renewcommand{\outer}{\msr{O}}
\newcommand{\outerprime}{\msr{O}'}
\newcommand{\region}{\msr{P}}
\newcommand{\Q}{\mathsf Q}
\newcommand{\lambert}{\mathsf W}

\newcommand{\de}{\mathrm{d}}

\newcommand{\oleq}[1]{\overset{\text{(#1)}}{\leq}}
\newcommand{\oeq}[1]{\overset{\text{(#1)}}{=}}
\newcommand{\ogeq}[1]{\overset{\text{(#1)}}{\geq}}
\newcommand{\ogeql}[2]{\overset{#1}{\underset{#2}{\gtreqless}}}

\newtheorem{mydef}{Definition}
\newtheorem{prop}{Proposition}
\newtheorem{theorem}{Theorem}
\newtheorem{lemma}{Lemma}
\newtheorem{example}{Example}
\newtheorem{corollary}{Corollary}
\newtheorem{remark}{Remark}

%%%%%%%%%%%%%%%%%%%%%%%%%%%%%%%%%%%%%%%%%%%%%%%%%%%%%%%%%%%%%%%%%%%%%%%

%%%%%%%%%%%%%%%%%%%%%%%%%%%%%%%%%%%%%%%%%%%%%%%%%%%%%%%%%%%%%%%%%%%%%%%%%%%%%%%%%%%%%%%%%%%%%%%%%%%%
%%%%%%%%%%%%%%%%%%%%%%%%
\title{Distance Spectrum of Fixed-Rate Raptor Codes with Linear Random Precoders}

\author{
Francisco L\'azaro, \IEEEmembership{Student Member, IEEE}, Enrico Paolini, \IEEEmembership{Member, IEEE},\\
Gianluigi Liva, \IEEEmembership{Senior Member, IEEE}, Gerhard Bauch, \IEEEmembership{Fellow, IEEE}
\thanks{Francisco L\'azaro and Gianluigi Liva are with the Institute of Communications and
Navigation of the German Aerospace Center (DLR), Muenchner Strasse 20, 82234
Wessling, Germany.
Email:\{\texttt{Francisco.LazaroBlasco}, \texttt{Gianluigi.Liva}\}\texttt{@dlr.de}.}
\thanks{Enrico Paolini is with the Department of Electrical, Electronic, and Information Engineering ``G. Marconi'', University of Bologna, via Venezia 52, 47521 Cesena (FC), Italy.
E-mail: \texttt{e.paolini@unibo.it}.}
\thanks{Gerhard Bauch is with the Institute for Telecommunication,  Hamburg University of Technology, Hamburg, Germany.
E-mail: \texttt{Bauch@tuhh.de}.}
\thanks{Corresponding Address:
Francisco L\'azaro, KN-SAN, DLR, Muenchner Strasse 20, 82234 Wessling, Germany. Tel: +49-8153 28-3211, Fax: +49-8153 28-2844, E-mail: \texttt{Francisco.LazaroBlasco@dlr.de}.}
\thanks{This work has been presented in part at the 2015 IEEE International Symposium on Information Theory, Hong Kong, China, June 2015.}
\thanks{This work has been accepted for publication in the special issue on `` Recent Advances in Capacity Approaching Codes'' in the Journal on Selected Areas in Communications, 2015.}
\thanks{\copyright 2015 IEEE. Personal use of this material is permitted. Permission
from IEEE must be obtained for all other uses, in any current or future media, including
reprinting /republishing this material for advertising or promotional purposes, creating new
collective works, for resale or redistribution to servers or lists, or reuse of any copyrighted
component of this work in other works}
} \maketitle
%%%%%%%%%%%%%%%%%%%%%%%%%%%%%%%%%%%%%%%%%%%%%%%%%%%%%%%%%%%%%%%%%%%%%%%%%%%%%%%%%%%%%%%%%%%%%%%%%%%%
%%%%%

\thispagestyle{empty}

%%%%%%%%%%%%%%%%%%%%%%%%%%%%%%%%%%%%%%%%%%%%%%%%%%%%%%%%%%%%%%%%%%%%%%%%%%%%%%%%%%%%%%%%%%%%%%%%%%%%
%%%%%%%%%%%%%%%%%%%%%%%%%

\begin{abstract}
Raptor code ensembles with linear random outer codes in a fixed-rate setting are considered. An expression for the average distance spectrum is derived and
this expression is used to obtain the asymptotic exponent of the weight distribution.
The asymptotic growth rate analysis is then exploited to develop a necessary and sufficient condition under which the fixed-rate Raptor code ensemble exhibits a strictly positive typical minimum distance. The condition involves the rate of the outer code, the rate of the inner fixed-rate \ac{LT} code and the LT code degree distribution. Additionally, it is shown that for ensembles fulfilling this condition, the minimum distance of a code randomly drawn from the ensemble has a linear growth with the block length. The analytical results can be used to make accurate predictions of the performance of finite length Raptor codes. These results are particularly useful for fixed-rate Raptor codes under \acl{ML} erasure decoding, whose performance is driven by their weight distribution.
\end{abstract}

%%%%%%%%%%%%%%%%%%%%%%%%%%%%%%%%%%%%%%%%%%%%%%%%%%%%%%%%%%%%%%%%%%%%%%%%%%%%%%%%%%%%%%%%%%%%%%%%%%%%%%%%%%

%\begin{keywords}

%%%%%%%%%%%%%%%%%%%%%%%%%%%%%%%%%%%%%%%%%%%%%%%%%%%%%%%%%%%%%%%%%%%%%%%
\begin{acronym}
\acro{WE}{weight enumerator}
\acro{WEF}{weight enumerator function}
\acro{IOWEF}{input output weight enumerator function}
\acro{IOWE}{input output weight enumerator}
\acro{LT}{Luby Transform}
\acro{BP}{belief propagation}
\acro{ML}{maximum likelihood}
\acro{MDS}{maximum distance separable}
\acro{LDPC}{low density parity check}
\acro{i.i.d.}{independent and identically distributed}
\acro{CER}{codeword error rate}
\acro{BEC}{Binary Erasure Channel}
\acro{ARQ}{automatic repeat request}

\acro{TCP}{transmission control protocol}
\acro{LDPC}{Low-density parity-check}

\end{acronym}

%%%%%%%%%%%%%%%%%%%%%%%%%%%%%%%%%%%%%%%%%%%%%%%%%%%%%%%%%%%%%%%%%%%%%%%

%\clearpage
%\tableofcontents
%\clearpage
\section{Introduction}\label{sec:Intro}
\IEEEoverridecommandlockouts

%\PARstart{E}{rasure}
Erasure
 channels, the first example of which was introduced in \cite{Elias55:2noisy}, have attracted an increasing  attention in the last decades. Originally regarded as purely theoretical channels, they turned out to be a very good abstraction model for the transmission of data over the Internet, where packets get lost randomly due to, for example, buffer overflows at intermediate routers. Erasure channels also find applications in wireless and satellite channels where deep fading events can cause the loss of one or multiple packets.

Traditionally, \ac{ARQ} mechanisms have been used in order to achieve reliable communication. A good example is the \ac{TCP} that is used for data transmission over the Internet.  \ac{ARQ} relies on feedback from the receiver and retransmissions and it is known to perform poorly when the delay between transmitter and receiver is high or when multiple receivers are present (reliable multicasting). An early work on erasure coding is \cite{Metzer84:retransmission}, where Reed-Solomon codes and (dense) linear random codes are proposed. However, those techniques become impractical due to their complexity already for small block lengths. More recently Tornado codes were proposed for transmission over erasure channels \cite{Studio3:PracticalLossRes, luby2001efficient}. Tornado codes have linear encoding and decoding complexities (under \acl{BP} decoding). However, the encoding and decoding complexities are proportional to their block lengths and not their dimension. Hence, they are not suitable for low rate applications such as reliable multicasting in which the transmitter needs to adapt its code rate to the user with the worst channel (highest erasure probability). \ac{LDPC} codes have also been proposed for use over erasure channels \cite{oswald2002capacity, miller04:bec, paolini2012maximum} and they have been proved to be practical in several scenarios even under \ac{ML} decoding. %However, also in this case the decoding complexity grows with the erasure rate of the channel.

Fountain codes \cite{byers02:fountain} are erasure codes potentially able to generate an endless amount of encoded symbols.  They find application in contexts where the channel erasure rate is not known a priori.
 The first class of practical fountain codes, \ac{LT} codes, was introduced in \cite{luby02:LT} together with an iterative \ac{BP} decoding algorithm that is efficient when the number of input symbols $k$ is large. One of the shortcomings of \ac{LT} codes is that in order to have a low probability of unsuccessful decoding, the encoding cost per output symbol has to be $\mathcal O \left(\ln(k)\right)$. Raptor codes were introduced in \cite{shokrollahi2001raptor}  \cite{shokrollahi06:raptor} as an evolution of \ac{LT} codes. They were also independently proposed in \cite{maymounkov2002online}, where they are referred to as online codes. Raptor codes consist of a serial concatenation of an outer code $\mathcal C$ (usually called precode) with an inner \ac{LT} code. The \ac{LT} code design can thus be relaxed requiring only the recovery of a fraction $1-\gamma$ of the input symbols with $\gamma$ small. This can be achieved with linear encoding complexity. The outer code is responsible for recovering the remaining fraction of input symbols, $\gamma$. If the outer code $\mathcal C$ is linear-time encodable, then the Raptor code has a linear encoding complexity, $\mathcal O\left( k \right)$, and therefore the overall encoding cost per output symbol is constant with respect to $k$. If \ac{BP} decoding is used, the decoding complexity is also linear in the dimension $k$ and not in the blocklegth $n$, as it is the case for LDPC and Tornado codes. This leads to a constant decoding cost per symbol, regardless  of the blocklength (i.e., of the rate). Furthermore, in \cite{shokrollahi06:raptor} it was shown that Raptor codes under \ac{BP} decoding are universally capacity-achieving on the binary erasure channel.
%Thus, they allow to achieve the capacity of a binary erasure channel even when the erasure probability is not known a priori.

Most of the works on \ac{LT} and Raptor codes consider \ac{BP} decoding which has a good performance for very large input blocks ($k$ at least in the order of a few tens of thousands symbols). Often in practice, smaller blocks are used. For example, for the Raptor codes standardized in \cite{MBMS12:raptor} and \cite{luby2007rfc} the recommended values of $k$ range from $1024$ to $8192$. For these input block lengths, the performance under \ac{BP} decoding degrades considerably. In this context, an efficient \ac{ML} decoding algorithm in the form of inactivation decoding \cite{shokrollahi2005systems} may be used in place of \ac{BP}. Some recent works have studied the decoding complexity of  Raptor and \ac{LT} codes under inactivation decoding \cite{Lazaro:ITW104,Lazaro:SCC15,mahdaviani2012raptor}.
In \cite{barg01:concat} lower bounds on the distance and error exponent are derived for a concatenated scheme with random outer code and a fixed inner code.
In \cite{ShokrollahiNow:2009} it is shown how the rank profile of the constraint matrix of a Raptor code depends on the rank profile of the outer code parity check matrix and the generator matrix of the LT code.
In \cite{Rahnavard:07} upper and lower bounds on the bit error probability of \ac{LT} and Raptor codes under \ac{ML} decoding are derived. The outer codes there considered in this work are picked from a linear ensemble in which the elements of the parity check matrix are independently set to one with a given probability $p<1/2$.
This work is extended in \cite{schotsch2011performance}, where upper and lower bounds on the codeword error probability of \ac{LT} codes under \ac{ML} decoding are developed. Another extension of this work is \cite{Schotsch:14}  where a pseudo upper bound on the performance of Raptor codes under \ac{ML} decoding is derived under the assumption that the number of erasures correctable by the outer code is small. Hence, this approximation holds only if the rate of the outer code is sufficiently high.
In \cite{wang:2015a} lower and upper bounds on the probability of successful decoding of \ac{LT} codes under \ac{ML} decoding as a function of the receiver overhead are derived, while corresponding bounds are developed in \cite{wang:2015} for Raptor codes. In \cite{tsung:15} finite length protograph-based Raptor-like LDPC codes are proposed for the AWGN channel.

Despite their rateless capability, Raptor codes represent an excellent solution also for fixed-rate communication schemes requiring powerful erasure correction capabilities with low decoding complexity. It is not surprising that Raptor codes are used in a fixed-rate setting by some existing communication systems (see, e.g., \cite{DVB-SH:raptor}). In this context, the performance under  \ac{ML} erasure decoding is determined by the distance properties of the fixed-rate Raptor code ensemble.

In contrast to \cite{ShokrollahiNow:2009, Schotsch:14, wang:2015}, in this work  we consider Raptor codes in a fixed-rate setting analyzing their distance properties.  In particular, we focus on the case where the outer code is picked from the linear random code ensemble. The choice of this ensemble is not arbitrary. The outer code used by the R10 Raptor code, the most widespread version of binary Raptor codes (see  \cite{MBMS12:raptor,luby2007rfc}),  is a concatenation of two systematic codes, the first being a high-rate regular \ac{LDPC} code and the second a pseudo-random code characterized by a dense parity check matrix. The outer codes of  R10 Raptor codes were designed to behave as codes drawn from the linear random ensemble in terms of rank properties, but allowing a fast algorithm for matrix-vector multiplication \cite{ShokrollahiNow:2009}. Thus,
 the ensemble we analyze may be seen as a simple model for practical Raptor codes with outer codes specifically designed to mimic the behavior of linear random codes. This model has the advantage to make the analytical investigation tractable. Moreover, although it is simple, the results obtained using this model allow predicting the behavior of binary Raptor codes in the standards rather accurately, as illustrated by simulation results in this paper.

 For the considered Raptor ensemble we develop a necessary and sufficient condition to guarantee a strictly positive normalized typical minimum distance, that involves the degree distribution of the inner fixed-rate \ac{LT} code, its rate, and the rate of the outer code. It identifies a positive normalized typical minimum distance region on the $(\ri,\ro)$ plane, where $\ri$ and $\ro$ are the inner and outer code rates. This can be used as an instrument for fixed-rate Raptor code desing. In particular, for a given overall rate $\rate$ of the fixed-rate Raptor ensemble, it allows to identify the smallest fraction of $\rate$ that has to be assigned to the outer code to obtain good distance properties. A necessary condition is also derived which, beyond the inner/outer code rates, depends on the average output degree only.
Finally we show how the analytical results presented in this paper may be used to predict the performance of finite length fixed-rate Raptor codes. This work extends the earlier conference paper \cite{lazaro:2015}. \footnote{ In this paper we provide full proofs of all the results developed in \cite{lazaro:2015}. More in detail, rigorous proofs of the growth rate expression (Theorem~2) and of the positive distance region (Theorem~3) are provided, together with both new results on the distance properties of the considered fixed-rate Raptor codes (Theorem~4 and Theorem~5) and performance curves of finite length codes obtained via software simulations.}

The rest of the paper is organized as follows. In Section \ref{sec:ensemble} we introduce the main definitions. Section~\ref{sec:dist} provides the derivation of the average weight distribution of the Raptor code ensemble considered and the associated growth rate.
Section~\ref{sec:rate_reg} provides necessary and sufficient conditions for a linear growth of the minimum distance with the block length (positive normalized typical minimum distance).
Numerical results are presented in Section~\ref{sec:results}. The conclusions follow in Section~\ref{sec:Conclusions}.

\section{Preliminaries}\label{sec:ensemble}
\begin{figure}
        \centering
        \includegraphics[width=0.9\columnwidth]{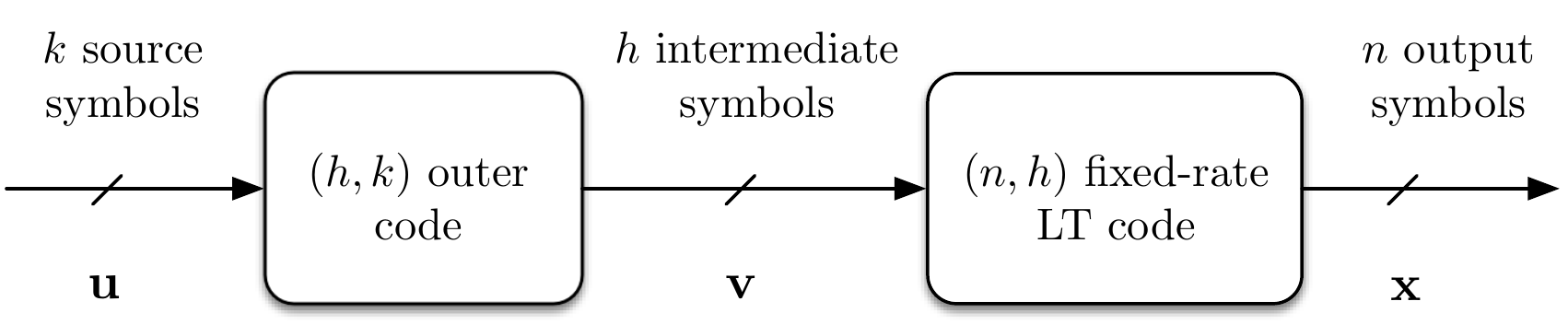}
        \caption{A Raptor code consists of a serial concatenation of an (outer) linear block code with an \ac{LT} code.}
        \label{fig:raptor}
\end{figure}

We consider fixed-rate Raptor code ensembles based on the encoder structure depicted in Figure \ref{fig:raptor}. The encoder is given by a serial concatenation of an $(h,k)$ outer code with an $(n,h)$ inner fixed-rate \ac{LT} code. We denote by $\vecu$ the outer encoder input, and by $\vecU$ the corresponding random vector. Similarly, $\vecv$ and $\vecx$ denote the input and the output of the fixed-rate \ac{LT} encoder, with $\vecV$ and $\vecX$ being the corresponding random vectors. The vectors $\vecu$, $\vecv$, and $\vecx$ are composed by $k$, $h$, and $n$ symbols respectively. The symbols of $\vecu$ are referred to as \emph{source} symbols, whereas the symbols of $\vecv$ and $\vecx$ are referred to as \emph{intermediate} and \emph{output} symbols, respectively.

We restrict ourselves to symbols belonging to $\field$. We denote by $\hw(\veca)$ the Hamming weight of a binary vector $\veca$. For a generic \ac{LT} output symbol $x_i$, $\deg (x_i)$ denotes the output symbol degree, i.e., the number of intermediate symbols that are added (in $\mathbb F_2$) to produce $x_i$.
%We will denote by $\ro=k/h$,  $\ri=h/n$, and $\rate=k/n=\ro \ri$ the rates of the outer code, the inner \ac{LT} code, and the Raptor code, respectively.
We will denote by $\ro=k/h$,  $\ri=h/n$, and $\rate=k/n=\ro \ri$ the rates of the outer, inner \ac{LT} codes. %and the Raptor code, respectively.

We consider the ensemble of Raptor codes $\ensemble(\oensemble,\Omega, \ri, \ro, n)$ obtained by a serial concatenation of an outer code in the $\left(\ri n,\ro\ri n\right)$ binary linear random block code ensemble $\oensemble$, with all possible realizations of an $\left(n,\ri n\right)$ fixed-rate \ac{LT} code with output degree distribution $\Omega= \{ \Omega_1, \Omega_2,\Omega_3, \ldots, \Omega_{\dmax}\}$, where $\Omega_i$ is the probability of having an output symbols of degree $i$. We also denote by $\bar \Omega$ the average output degree, $
\bar \Omega = \sum_i i\Omega_i$.

Picking randomly one code in the ensemble $\ensemble(\oensemble,\Omega, \ri, \ro, n)$ is performed by randomly drawing the parity-check matrix of the linear random outer code and the low density generator matrix of the fixed-rate \ac{LT} encoder. The parity-check matrix of the outer code is obtained by drawing $(h-k)h$ \ac{i.i.d.} Bernoulli uniform random variables. The generator matrix of the fixed-rate \ac{LT} encoder is generated by independently drawing $n$ degrees $i$ according to the probability mass function (p.m.f.) $\Omega$ and, for each such degree $i$, by choosing uniformly at random $i$ distinct symbols out of the $h$ intermediate ones.

We make use of the notion of exponential equivalence \cite{CoverThomasBook}. Two real-valued positive  sequences $a(n)$ and $b(n)$ are said to be exponentially equivalent, writing $a(n)~\doteq~b(n)$, when
%
%\vspace{-0.25cm}
\begin{equation}\label{eq:asymp_eq}
\lim_{n \to \infty} \frac{1}{n} \log_2 \frac{a(n)}{b(n)}=0.
\end{equation}
If $a(n)$ and $b(n)$ are exponentially equivalent, then \begin{align}
\lim_{n \to \infty} \frac{1}{n} \log_2 a(n) = \lim_{n \to \infty} \frac{1}{n}  \log_2 b(n).
\end{align}
Given two pairs of reals $(x_1,y_1)$ and $(x_2,y_2)$, we write $(x_1,y_1) \succeq (x_2,y_2)$ if $x_1 \geq x_2$ and $y_1 \geq y_2$.

\section{Distance Spectrum of Fixed-Rate Raptor Code Ensembles}\label{sec:dist}
In this section we characterize the expected \ac{WE} of a fixed-rate Raptor code picked randomly in the ensemble $\ensemble(\oensemble,\Omega, \ri, \ro, n)$. An expression for the expected \ac{WE} is first obtained. Then, the asymptotic exponent of the \ac{WE} is analyzed.

\begin{theorem}\label{theorem:we}
Let $A_\d$ be the expected multiplicity of codewords of weight $\d$ for a code picked randomly in the ensemble $\ensemble(\oensemble,\Omega, \ri, \ro, n)$. For $d\geq1$ we have
\begin{align}\label{eq:WEF_Raptor}
A_\d = \binom {n}{\d} 2^{-h (1-\ro)} \sum_{\l=1}^h \binom{h}{\l}   \pl^\d (1-\pl)^{n-\d}
\end{align}
where
\begin{align}\label{eq:pl_finite}
\pl &= \sum_{j=1}^{\dmax} \Omega_j \sum_{\substack{i=\max(1,\l+j-h)\\ i~\mathrm{odd}}}^{ \min (\l,j)} \frac{ \binom{j}{i} \binom{h-j}{\l-i} } { \binom{h}{\l}} \\
         &= \sum_{j=1}^{\dmax} \Omega_j \sum_{\substack{i=\max(1,\l+j-h)\\ i~\mathrm{odd}}}^{ \min (\l,j)} \frac{ \binom{\l}{i} \binom{h-\l}{j-i} } { \binom{h}{j}} \, .
\end{align}
\end{theorem}
\begin{IEEEproof}
For a serially concatenated code we have
\begin{align}
A_\d = \sum_{\l=1}^{h} \frac{\weo_{\l} \wei_{\l,\d}}{ \binom {h} {\l}}
\label{eq:we_serial}
\end{align}
where $\weo_{\l}$ is the average \ac{WE} of the outer code, and $\wei_{\l,\d}$ is the average \ac{IOWE} of the inner fixed-rate \ac{LT} code. The average \ac{WE} of an $(h,k)$ linear random code is known to be \cite{Gallager63}
\begin{align}
\weo_{\l} = \binom{h}{l} 2^{-h (1-\ro)}.
\label{eq:wef_random}
\end{align}
We now focus on the average \ac{IOWE} of the fixed-rate \ac{LT} code.  Let us denote by $l$ the Hamming weight of the input word to the \ac{LT} encoder and let us denote by $\pjl$ the probability that any of the $n$ output bits of the \ac{LT} encoder takes the value $1$ given that the Hamming weight of the intermediate word  is $\l$ and the degree of the \ac{LT} code output symbol is $j$, i.e.,
\[
\pjl:=\Pr\{X_i=1|\hw(\vecV)=\l,\deg(X_i)=j\}
\]
for any $i\in \{1,\dots,n\}$. This probability may be expressed as
\begin{align}
\pjl =
\sum_{\substack{i=\max(1,\l+j-h)\\ i~\textrm{odd}}} ^{ \min (\l,j)} \frac{ \binom{j}{i} \binom{h-j}{\l-i} } { \binom{h}{\l} } \,= \sum_{\substack{i=\max(1,\l+j-h)\\ i~\textrm{odd}}} ^{ \min (\l,j)} \frac{ \binom{\l}{i} \binom{h-\l}{j-i} } { \binom{h}{j}}
\label{eq:p_j_l}
\end{align}
Removing the conditioning on $j$ we obtain $\pl$, the probability of any of the $n$ output bits of the fixed-rate \ac{LT} encoder taking value $1$ given a Hamming weight $l$ for the intermediate word, i.e.,
\[
\pl:=\Pr\{X_i=1|\hw(\vecV)=l\}
\]
for any $i\in \{1,\dots,n\}$.
We have
\begin{align}
\pl = \sum_{j=1}^{\dmax} \Omega_j \pjl.
\label{eq:p_l}
\end{align}
Since the output bits are generated by the \ac{LT} encoder independently of each other, the Hamming weight of the \ac{LT} codeword conditioned to an intermediate word of weight $l$ is a binomially distributed random variable with parameters $n$ and $\pl$. Hence, we may write
\begin{align}
\Pr\{\hw(\vecX) = \d | \hw(\vecV) = \l\}
=\binom {n}{\d} \pl^\d (1-\pl)^{n-\d}.\label{eq:distr_weight_LT}
\end{align}
The average \ac{IOWE} of a \ac{LT} code may now be easily calculated multiplying \eqref{eq:distr_weight_LT} by the number of weight-$\l$ intermediate words, yielding
\begin{align}
\wei_{\l,\d}= \binom {h}{\l} \binom {n}{\d} \pl^\d (1-\pl)^{n-\d}.
\label{eq:iowef_lt}
\end{align}
Making use of \eqref{eq:we_serial}, \eqref{eq:wef_random} and \eqref{eq:iowef_lt}, we obtain \eqref{eq:WEF_Raptor}.
\end{IEEEproof}
\begin{remark}
As opposed to $A_d$ with $d \geq 1$, whose expression is given by (1), the expected number of codewords of weight $0$, $A_0$, is given by
\begin{align*}
A_0 &= 1 + \sum_{l=1}^h \frac{\weo_{\l} \wei_{\l,0}}{ \binom {h} {\l}} \\
       &= 1 + 2^{-n \ri (1 - \ro) } \sum_{l=1}^h \binom {h} {\l} (1 - \pl)^n \, .
\end{align*}
An expected number of weight-$0$ codewords larger than one is related to the fact that we have a nonzero probability that the $h \times n$ generator matrix of the fixed-rate LT code is not full-rank. This matrix, in fact, is generated ``online'' in the standard way for LT encoding, i.e., by drawing $n$ i.i.d. discrete random variables with p.m.f. $\Omega$, representing the weights of the $n$ columns. For each such column, the corresponding `$1$' entries are placed in random positions. It will be shown in Section~\ref{sec:rate_reg}, Theorem~\ref{theorem:zero_codeword}, that if the $(\ri,\ro)$ pair belongs to the region there called ``positive normalized typical minimum distance region'', the expected number $A_0$ of zero weight codewords approaches $1$ (exponentially) as $n$ increases.
\end{remark}

%\subsection{Growth Rate of Fixed-Rate Raptor Code Ensembles}\label{sec:growth}

Next we compute the asymptotic exponent (growth rate) of the weight distribution for the ensemble $\msr{C}_{\infty}(\oensemble,\Omega, \ri, \ro)$, that is the ensemble $\msr{C}(\oensemble,\Omega, \ri, \ro, n)$ in the limit where $n$ tends to infinity for constant $\ri$ and $\ro$.
Hereafter, we denote the normalized output weight of the Raptor encoder by $\nd = \d/n$ and the normalized output weight of the outer code (input weight to the \ac{LT} encoder) by $\nl = \l/h$. The growth rate is defined as
\begin{align}
\G(\nd) = \lim_{n \to \infty} \frac{1}{n} \log_2 \we_{\nd n} \, . \label{eq:growth_rate_def}
\end{align}

\begin{theorem}\label{theorem:growth_rate}
The asymptotic exponent of the weight distribution of the fixed-rate Raptor code ensemble $\msr{C}_{\infty}(\oensemble,\Omega, \ri, \ro)$ is given by
\begin{align}\label{eq:growth_rate}
\G(\nd) = \Hb(\nd) - \ri  (1-\ro) +  \fmax(\nd)
\end{align}
where
\begin{align}\label{eq:max}
\fmax(\nd) := \max_{ \nl \in \mathscr D_{\nl}} \f(\nd, \nl),
\end{align}
being $\f(\nd, \nl)$ and $\mathscr D_{\nl}$ defined as follows,
\begin{align}\label{eq:f}
\f(\nd, \nl) := \ri \Hb(\nl) + \nd \log_2 \npnl + (1- \nd) \log_2 \left(1 - \npnl\right) \, ,
\end{align}
\begin{align}
\mathscr D_{\nl} = \left\{ \begin{array}{cl} (0,1) & \textrm{if  } \Omega_j = 0 \textrm{  for all even } j\\
(0,1] & \textrm{otherwise} \, , \end{array} \right.
\end{align}
with $\npnl$ defined as
\begin{align}
\npnl := \frac{1}{2} \sum_{j=1}^{\dmax} \Omega_j  \left[  1-\left( 1-2\nl\right)^j \right].
\label{eq_npnl}
\end{align}
\end{theorem}
\begin{IEEEproof}
Let us define $\mathbb N^*_h = \{1,2,\dots,h\}$. From \eqref{eq:WEF_Raptor} we have
\begin{align} \label{eq:proof_G}
                                              & \frac{1}{n} \log_2 A_{\delta n} \notag \\
%                                              &= \frac{1}{n} \log_2 \left( {n \choose d} 2^{-n  \ri (1- \ro)} \sum_{l=1}^h {h \choose l} \pl^d (1-\pl)^{n-d} \right) \\
                                              &=\frac{1}{n} \log_2 {n \choose \delta n} -  \ri (1- \ro) + \frac{1}{n} \log_2  \sum_{l=1}^h {h \choose l} \pl^d (1-\pl)^{n-d}  \notag \\
                                              &\stackrel{\mathrm{(a)}}{\leq} \Hb(\delta) -\frac{1}{2n} \log_2 \left(2 \pi n \delta (1-\delta)\right) -  \ri (1- \ro) \\
                                              &+ \frac{1}{n} \log_2 \sum_{l=1}^h {h \choose l} \pl^d (1-\pl)^{n-d} \notag \\
                                              &\stackrel{\mathrm{(b)}}{\leq} \Hb(\delta) -\frac{1}{2n} \log_2 \left(2 \pi n \delta (1-\delta)\right) -  \ri (1- \ro) \\
                                              &+ \frac{1}{n}\log_2 ( \ri n) %\notag \\
                                              + \frac{1}{n} \log_2 \max_{l \in \mathbb N^*_{h-1}} \left\{ {h \choose l} \pl^d (1-\pl)^{n-d} \right\} \notag \\
                                              &\stackrel{\mathrm{(c)}}{\leq} \Hb(\delta) -\frac{1}{2n} \log_2 (2 \pi n \delta (1-\delta)) -  \ri (1- \ro) + \frac{1}{n}\log_2 ( \ri n) \notag \\
                                             &+ \max_{l \in \mathbb N^*_{h-1}} \left\{  \ri \Hb\left(\frac{l}{h}\right) - \frac{1}{2n}\log_2\left(2 \pi  \ri n\frac{l}{h}\left(1-\frac{l}{h}\right)\right) \right.\\
                                             & + \delta \log_2 \pl + (1-\delta) \log_2(1-\pl) \Big\} \notag \\
                                             &= \Hb(\delta) -\frac{1}{2n} \log_2 (2 \pi n \delta (1-\delta)) -  \ri (1- \ro) + \frac{1}{n}\log_2 ( \ri n) \notag \\
                                             &+ \max_{\nl \in \left\{\frac{1}{\ri n},\dots,\frac{\ri n -1}{\ri n}\right\}} \left\{  \ri \Hb\left(\nl\right) - \frac{1}{2n}\log_2\left(2 \pi  \ri n \nl \left(1- \nl\right)\right) \right.\\
                                             &+ \delta \log_2 p_{\ri n \lambda} + (1-\delta) \log_2(1-p_{\ri n \lambda}) \Big\}
\end{align}
Inequality $\mathrm{(a)}$ follows from the well-known tight bound \cite{Gallager63}
\begin{align}
{n \choose \sigma n} \leq \frac{2^{n \Hb(\sigma)}}{\sqrt{2 \pi n \sigma (1-\sigma)}}, \qquad 0<\sigma<1
\label{eq:gallagher_upper}
\end{align}
while $\mathrm{(b)}$ from
\begin{align}
\sum_{l=1}^h {h \choose l} \pl^d (1-\pl)^{n-d} \leq h \max_{l \in \mathbb N^*_h} {h \choose l} \pl^d (1-\pl)^{n-d}
\label{eq:proof_G_summ}
\end{align}
and from the fact that the maximum cannot be taken for $l=h$ for large enough $h=\ri n$, hence for large enough $n$ (as shown next). Inequality $\mathrm{(c)}$ is due again to \eqref{eq:gallagher_upper}, to $\log_2(\cdot)$ being a monotonically increasing function, and to $1/n$ being a scaling factor not altering the result of the maximization with respect to $l$.

That the maximum is not taken for $l=h$, for large enough $h$, may be proved as follows. By direct calculation of \eqref{eq:p_l} for $l=h$ and $l=h-1$ it is easy to show that we have
\begin{align*}
\p_h = \sum_{\substack{j=1\\ j~\textrm{odd}}}^{d_{\max}} \Omega_j ~~ \mathrm{and}
\qquad \p_{h-1} = \sum_{\substack{j=1\\ j~\textrm{odd}}}^{d_{\max}} \frac{h-j}{h} \Omega_j + \sum_{\substack{j=1\\ j~\textrm{even}}}^{d_{\max}} \frac{j}{h} \Omega_j \, .
\end{align*}
Since $\p_{h-1}/\p_h \rightarrow 1$ for increasing $h$, there exists $h_0(\Omega)$ such that
\begin{align}
h\, \p_{h-1}^d (1-\p_{h-1})^{n-d} > \p_h^d (1-\p_h)^{n-d}
\end{align}
for all $h>h_0(\Omega)$. Hence, for all such values of $h$ the maximum cannot be taken at $l=h$.

Next, by defining
\begin{align}
\hat{\lambda}_n &= \mathop{\mathrm{argmax}}_{\lambda \in \left\{\frac{1}{\ri n}, \frac{2}{\ri n}, \dots, \frac{\ri n-1}{\ri n} \right\} } \Big\{   \ri H_b(\lambda)  - \frac{1}{2n}\log_2(2 \pi  \ri n\lambda(1-\lambda)) \\
&  + \delta \log_2 \p_{\ri n \lambda} + (1-\delta) \log_2(1-\p_{\ri n \lambda}) \Big\}
\label{eq:lambda_hat}
\end{align}
the right-hand side of \eqref{eq:proof_G} may be recast as
\begin{align}
                                             &\Hb(\delta) -\frac{1}{2n} \log_2 \left(2 \pi n \delta (1-\delta)\right) -  \ri (1- \ro) + \frac{1}{n}\log_2 ( \ri n) \notag \\
                                             & +  \ri \Hb(\hat{\lambda}_n) - \frac{1}{{2n}}\log_2(2 \pi  \ri n\hat{\lambda}_n(1-\hat{\lambda}_n)) \\
                                             &+ \delta \log_2 p_{\ri n \hat{\lambda}_n} + (1-\delta) \log_2(1-p_{\ri n \hat{\lambda}_n}) \, .
\end{align}
The two terms $\frac{1}{2n} \log_2 (2 \pi n \delta (1-\delta))$ and $\frac{1}{n} \log_2 (\ri n)$ in the last expression converge to zero as $n \rightarrow \infty$. Moreover, also the term $\frac{1}{{2n}}\log_2 (2 \pi  \ri n\hat{\nl}_n(1-\hat{\nl}_n))$ converges to zero regardless of the behavior of the sequence $\hat{\lambda}_n$. In fact, it is easy to check that the term $\frac{1}{{2n}}\log_2(2 \pi  \ri n\hat{\lambda}_n(1-\hat{\lambda}_n))$ converges to zero in the limiting cases $\hat{\lambda}_n=\frac{1}{\ri n}$ $\forall n$ and $\hat{\lambda}_n = \frac{\ri n-1}{\ri n}$ $\forall n$, so it does in all other cases. %The above three terms, therefore, are asymptotically vanishing so that in the limit as $n \rightarrow \infty$ the inequalities $\mathrm{(a)}$, $\mathrm{(b)}$, and $\mathrm{(c)}$ hold with equality.

Developing the right hand side of \eqref{eq:proof_G} further, for large enough $n$, we have
\begin{align}\label{eq:proof_G_2}
                                             & \Hb(\delta) -\frac{1}{2n} \log_2 (2 \pi n \delta (1-\delta)) -  \ri (1- \ro) + \frac{1}{n}\log_2 ( \ri n) \notag \\
                                             &+ \max_{\nl \in \left\{\frac{1}{\ri n},\dots,\frac{\ri n -1}{\ri n}\right\}} \left\{  \ri \Hb\left(\nl\right) - \frac{1}{2n}\log_2\left(2 \pi  \ri n \nl \left(1- \nl\right)\right) \right.\\
                                             &+ \delta \log_2 p_{\ri n \lambda} + (1-\delta) \log_2(1-p_{\ri n \lambda}) \bigg\} \notag \\
& \stackrel{\mathrm{(d)}}{\leq} \Hb(\delta) -\frac{1}{2n} \log_2 (2 \pi n \delta (1-\delta)) -  \ri (1- \ro) + \frac{1}{n}\log_2 ( \ri n) \notag \\
& + \sup_{\nl \in \mathbb Q \cap (0,1)} \bigg\{  \ri \Hb\left(\nl\right) - \frac{1}{2n}\log_2\left(2 \pi  \ri n \nl \left(1- \nl\right)\right) \\
&+ \delta \log_2 \left(\npnl + \frac{K}{n} \right) + (1-\delta) \log_2 \left(1-\npnl + \frac{K}{n} \right) \bigg\} \notag \\
& \stackrel{\mathrm{(e)}}{=} \Hb(\delta) -\frac{1}{2n} \log_2 (2 \pi n \delta (1-\delta)) -  \ri (1- \ro) + \frac{1}{n}\log_2 ( \ri n) \notag \\
& + \sup_{\nl \in (0,1)} \bigg\{  \ri \Hb\left(\nl\right) - \frac{1}{2n}\log_2\left(2 \pi  \ri n \nl \left(1- \nl\right)\right) \\
&+ \delta \log_2 \left(\npnl + \frac{K}{n} \right) + (1-\delta) \log_2 \left(1-\npnl + \frac{K}{n} \right) \bigg\} \\
&:= \Gamma_n (\nd).
\end{align}
where $\mathbb Q$ is the set of rational numbers. Inequality $\mathrm{(d)}$ follows from the fact that, as it can be shown, $|\npnl - p_{\ri n \lambda}|<K/n$ (uniformly in $\lambda$) for large enough $n$ and from the fact that the supremum over $\mathbb Q \cap (0,1)$ upper bounds the maximum over the finite set $\left\{ \frac{1}{\ri n}, \dots, \frac{\ri n - 1}{\ri n} \right\}$. Equality $\mathrm{(e)}$ is due to the density of $\mathbb Q$. In equality $\mathrm{(e)}$, the function of $\nl$ being maximized is regarded as a function over the real interval $(0,1)$ (i.e., $\lambda$ is regarded as a real parameter).

The upper bound \eqref{eq:proof_G_2} on $\frac{1}{n} \log_2 A_{\nd n}$ is valid for any finite but large enough $n$. If we now let $n$ tend to infinity, all inequalities $\mathrm{(a)}$--$\mathrm{(d)}$ are satisfied with equality. In particular: for $\mathrm{(a)}$ this follows from the well-known exponential equivalence ${n \choose \nd n} \doteq 2^{n \Hb(\nd)}$; for $\mathrm{(b)}$ from the exponential equivalence $\sum_l 2^{n f(l)} \doteq \max_l 2^{n f(l)}$; for $\mathrm{(c)}$ from ${\ri n \choose \hat{\nl}_n \ri n} \doteq 2^{n \Hb(\hat{\nl}_n)}$ (due to $\frac{1}{{2n}}\log_2 (2 \pi  \ri n\hat{\lambda}_n(1-\hat{\lambda}_n))$ vanishing for large $n$); for $\mathrm{(d)}$ from the fact that, asymptotically in $n$, applying the definition of limit we can show that the maximum over the set $\left\{\frac{1}{\ri n}, \dots, \frac{\ri n -1}{\ri n} \right\}$ upper bounds the supremum over $\mathbb Q \cap (0,1)$ (while at the same time being upper bounded by it for any $n$). The expression of $\npnl$ is obtained by assuming $n$ tending to $\infty$ using the expression of $\pl$. Alternatively, the same expression is obtained by assuming $n$ tending to $\infty$ and letting an output symbol of degree $i$ choose its $i$ neighbors \emph{with} replacement.

By letting $n$ tend to infinity and by cancelling all vanishing terms, we finally obtain the statement. Note that we can replace the supremum by a maximum over $\mathscr D_{\nl}$ as this maximum is always well-defined.\footnote{In fact, for any $\nd \in [0,1]$ the function $\f(\nd, \nl)$ diverges to $-\infty$ as $\nl \rightarrow 0^+$. Moreover, it diverges to $-\infty$ as $\nl \rightarrow 1^-$ if $\Omega_j=0$ for all even $j$ and converges as $\nl \rightarrow 1^-$ otherwise. Finally, for all $\delta \in [0,1]$ it is continuous for all $\nl \in \mathscr D_{\nl}$.}
\end{IEEEproof}%%

The next two lemmas, which will be useful in the sequel, characterize the derivative of the growth rate function. For the sake of clarity, we use the notation $\np(\nl)$ instead of $\np_{\nl}$.

\begin{lemma} \label{lemma:growth_rate_derivative}
The derivative of the growth rate of the weight distribution of a fixed-rate Raptor code ensemble $\msr{C}_{\infty}(\oensemble,\Omega, \ri, \ro)$ is given by
\begin{equation}
G'(\nd) = \log_2 \frac{1-\nd}{\nd} + \log_2 \frac{\np(\nlo)}{1-\np(\nlo)}  \,  \nonumber
\end{equation}
where
\begin{align}\label{eq:lo_def}
\nlo(\nd) := \argmax_{\nl \in D_{\nl}} \left\{ \f(\nd, \nl) \right\} \, .
\end{align}
\end{lemma}
\begin{IEEEproof}
Let us rewrite the expression of $G(\nd)$ in \eqref{eq:growth_rate} as ${\G(\nd)=\Hb(\nd) - \ri(1-\ro) + \f(\nd,\nlo(\nd))}$. We must have
\begin{equation}\label{eq:critical_point}
\frac{\partial \f}{\partial \nl} (\nd, \nlo) = 0 \, .
\end{equation}
Taking the derivative with respect to $\nd$, after elementary algebraic manipulation we obtain
\begin{align*}
G'(\nd) &= \log_2\frac{1-\nd}{\nd} + \log_2 \frac{\np(\nlo)}{1-\np(\nlo)} + \frac{\partial \f}{\partial \nl} (\nd, \nlo) \, \frac{\mathrm d \nlo}{\mathrm d \nd}
\end{align*}
which, applying \eqref{eq:critical_point}, yields the statement.
\end{IEEEproof}
\begin{lemma}\label{corollary:der}
For all $0< \nd < 1/2$, the derivative of the growth rate of the weight distribution of a fixed-rate Raptor code ensemble $\msr{C}_{\infty}(\oensemble,\Omega, \ri, \ro)$ fulfills
\[
G'(\nd)>0.
\]
\end{lemma}
\begin{IEEEproof}
By imposing $G'(\nd)=0$, from Lemma~\ref{lemma:growth_rate_derivative} we obtain
$$
\frac{1-\nd}{\nd} = \frac{1-\varrho(\lambda_0)}{\varrho(\lambda_0)}
$$
which implies $\nd = \varrho(\lambda_0)$ since the function $(1-x)/x$ is monotonically decreasing for $x \in (0,1)$. Next, due to the definition of $\lambda_0$ in \eqref{eq:lo_def} we know that the partial derivative $\partial \mathsf f(\nd,\lambda) / \partial \lambda$ must be zero when calculated for $\lambda=\lambda_0$. The expression of this partial derivative is
$$
\frac{\partial \mathsf f}{\partial \lambda}(\nd,\lambda) = \mathsf r_{\mathsf i} \log_2 \frac{1 - \lambda}{\lambda} + \frac{\varrho'(\lambda)}{\log 2} \cdot \frac{\nd - \varrho(\lambda)}{\varrho(\lambda)(1-\varrho(\lambda))} \,
$$
so we obtain
$$
\mathsf r_{\mathsf i} \log_2 \frac{1 - \lambda_0}{\lambda_0} + \frac{\varrho'(\lambda_0)}{\log 2} \cdot \frac{\nd - \varrho(\lambda_0)}{\varrho(\lambda_0)(1-\varrho(\lambda_0))} = 0 \, .
$$
As shown above, for any $\nd$ such that $G'(\nd)=0$ we have $\nd=\varrho(\lambda_0)$. Substituting in the latter equation we obtain $\lambda_0=1/2$ which implies $\nd=\varrho(1/2)=1/2$. Therefore, the only value of $\nd$ such that $G'(\nd)=0$ is $\nd=1/2$. Due to continuity of $G'(\nd)$ and to the fact that $G'(\nd) \rightarrow +\infty$ as $\nd \rightarrow 0^+$ (as shown in Subsection~B of Appendix \ref{sec:proof_inversion}) we conclude that $G'(\nd)>0$ for all $0 < \nd < 1/2$.
\end{IEEEproof}

\begin{mydef}The normalized typical minimum distance of an ensemble $\msr{C}_{\infty}(\oensemble,\Omega, \ri, \ro)$  is the real number
%
%\begin{align*}
%\dmint := \inf \{ \nd>0 : \G(\nd) > 0 \}
%\end{align*}
\begin{align*}
\dmint := \begin{cases}
0 & \text{if } \lim_{\nd \to 0^+} \G(\nd) \geq 0 \\
\inf \{ \nd>0 : \G(\nd) > 0 \} & \text{otherwise.}
\end{cases}
\end{align*}
\end{mydef}
\begin{example}
Fig.~\ref{fig:growth} shows $\G(\nd)$ for the ensemble $\msr{C}_{\infty}(\oensemble,\Omega^{(1)}, \ri, \ro)$, where $\Omega^{(1)}$ is the output degree distribution used in the standards \cite{MBMS12:raptor}, \cite{luby2007rfc} (see details in Table~\ref{table:dist}) and $\ro=0.99$ for three different  $\ri$ values. The growth rate, $\G(\nd)$, of a linear random code ensemble with rate $\rate=0.99$ is also shown. It can be observed how the curve for $\ri = 0.95$ does not  cross the $x$-axis, the curve for $\ri = 0.88$ has $\dmint=0$ and the curve for $\ri=0.8$ has $\dmint=0.0005$.
\begin{figure}[!t]
\begin{center}
\includegraphics[width=\figw]{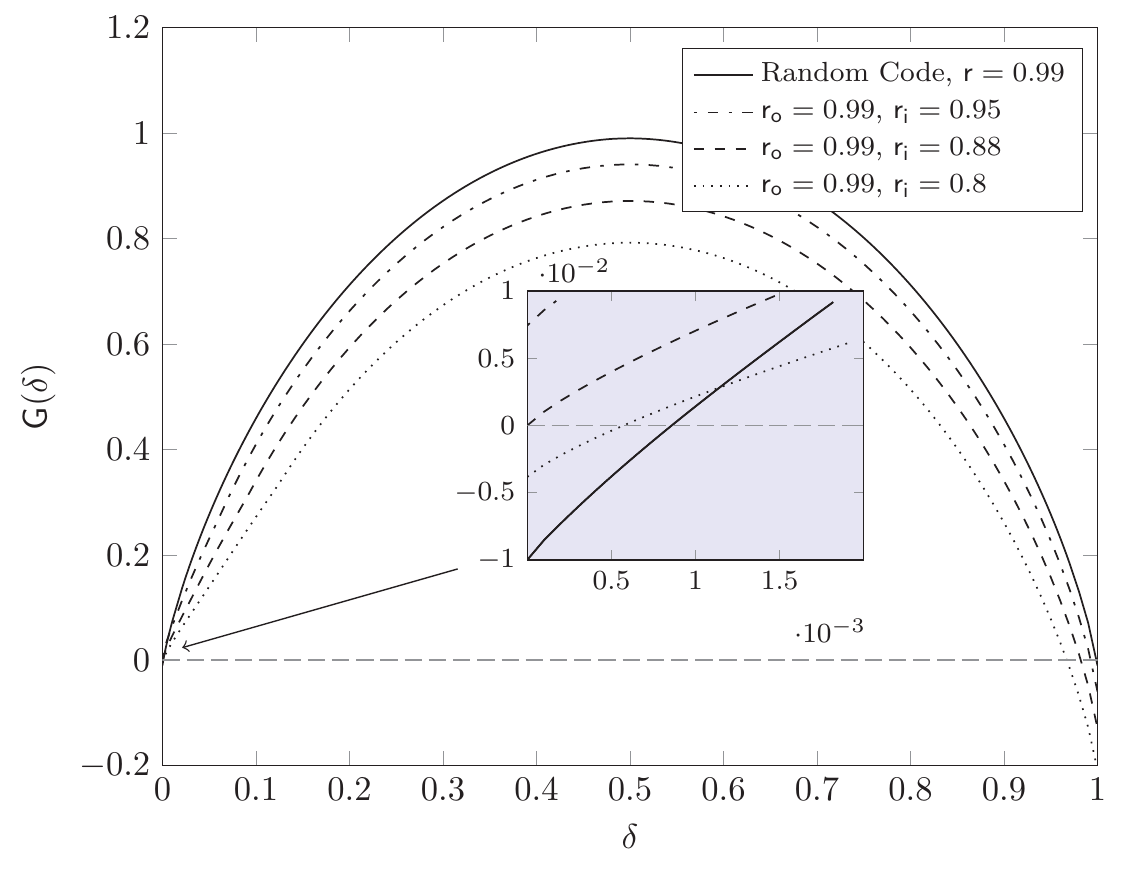}
\centering \caption{Growth rate vs. normalized output weight $\nd$. The continuous line shows the growth rate of a linear random code with rate $\rate=0.99$. The dot-dashed, dashed, and dotted lines show the growth rates $\G(\nd)$ of the ensemble $\msr{C}_{\infty}(\oensemble,\Omega^{(1)}, \ri, \ro=0.99)$ for $\ri=0.95$, $0.88$ and $0.8$, respectively.}
\label{fig:growth}
\end{center}
\end{figure}
\end{example}
\begin{example}
Fig.~\ref{fig:gilbert} shows the overall rate $\rate$ of the Raptor code ensemble $\msr{C}_{\infty}(\oensemble,\Omega^{(1)}, \ri~=~\rate/\ro, \ro)$ versus the normalized typical minimum distance $\dmint$. It can be observed how, for constant overall rate $\rate$, $\dmint$ increases as the outer code rate $\ro$ decreases. It also can be observed how decreasing $\ro$ allows to get closer to the asymptotic Gilbert-Varshamov bound.
\begin{figure}[h!]
\begin{center}
\includegraphics[width=\figw]{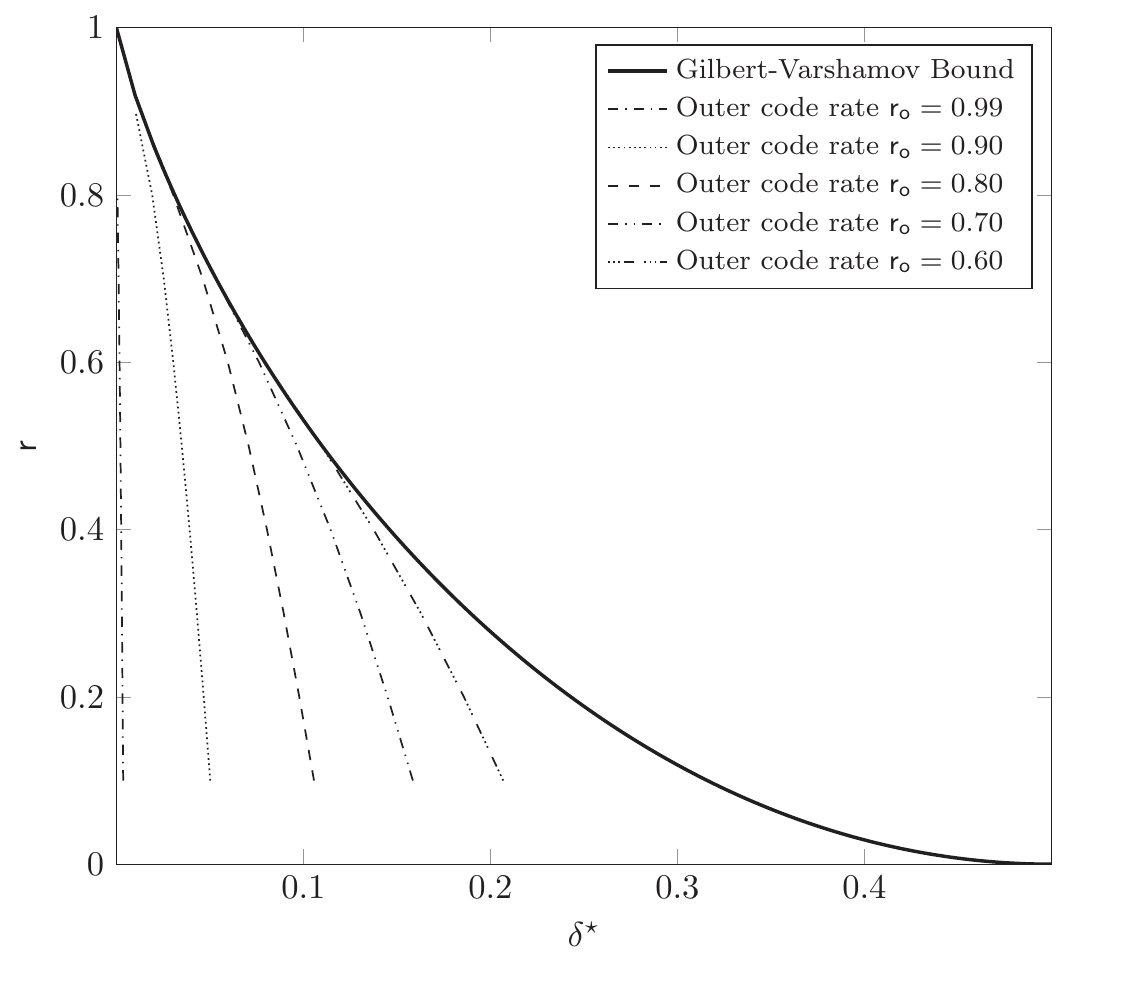}
\centering \caption{Overall rate $\rate$  vs. the normalized typical minimum distance $\dmint$. The continuous line represents the asymptotic Gilbert-Varshamov bound. The markers represent Raptor codes ensembles $\msr{C}_{\infty}(\oensemble,\Omega^{(2)}, \ri=\rate/ \ro, \ro)$ with different outer code rates, $\ro$.}
\label{fig:gilbert}
\end{center}
\end{figure}
\end{example}

\section{Typical Distance Rate Regions}
\label{sec:rate_reg}
In this section we aim at determining under which conditions the ensemble $\msr{C}_{\infty}(\oensemble,\Omega,\ri,\ro)$ exhibits good normalized typical distance properties. More specifically, given a distribution $\Omega$ and an overall rate $\rate$, we are interested in the allocation of the rate between the outer code and the fixed-rate LT code to achieve a strictly positive normalized typical minimum distance.
\begin{mydef}[Positive normalized typical minimum distance region] We define the \emph{positive} normalized typical minimum distance region of an ensemble $\msr{C}_{\infty}(\oensemble,\Omega,\ri,\ro)$ as the set $\region$ of code rate pairs $\left( \ri, \ro \right)$ for which the ensemble possesses a positive normalized typical minimum distance. Formally:
\begin{align}
\region:=\left\{(\ri,\ro) \succeq (0,0) | \dmint(\Omega, \ri,\ro)> 0 \right\}
\nonumber
\end{align}
where we have used the notation $\dmint= \dmint(\Omega, \ri,\ro)$ to emphasize the dependence on $\Omega$, $\ri$ and $\ro$.
\end{mydef}
The positive normalized typical distance region for an LT output degree distribution $\Omega$ is developed in the following theorem.
\begin{theorem} The region  $\region$ is given by
\label{theorem_inner}
\begin{align}
\region &:=\left\{\left( \ri, \ro \right) \succeq (0,0) | \ri (1-\ro) \right.\\
& \quad  \quad \quad> \max_{ \nl \in \mathscr D_{\nl}} \left\{\ri \Hb(\nl) + \log_2 \left(1 - \npnl\right)\right\}
\bigg\}\, .
\label{eq:theorem_region}
\end{align}
\end{theorem}
\begin{IEEEproof}
See Appendix~\ref{sec:proof_inversion}.
\end{IEEEproof}
The next two theorems characterize the distance properties of a fixed-rate Raptor code with linear random outer code picked randomly in the ensemble $\ensemble(\oensemble,\Omega, \ri, \ro, n)$ with $(\ri, \ro)$ belonging to $\region$.
\begin{theorem} \label{theorem:min_dist}
Let the random variable $\D$ be the minimum nonzero Hamming weight in the code book of a fixed-rate Raptor code picked randomly in an ensemble $\ensemble(\oensemble,\Omega, \ri, \ro, n)$. If $(\ri, \ro) \in \region$ then
\begin{align*}
\lim_{n \rightarrow \infty} \Pr \{ \D \leq \nd n \} = 0
\end{align*}
exponentially in $n$, for all $0 < \nd < \dmint$.
\end{theorem}
\begin{IEEEproof}
It is well known that this probability can be upper bounded via union bound as
\begin{align}
\label{eq:pr_dmin}
\Pr\{ \D \leq \nd n\} & \leq   \sum_{w=1}^{\nd n} A_w.
\end{align}
We will start by proving that the sequence $A_{\d}$ is non-decreasing for $\d < n/2$ and sufficiently large $n$.  As $n \rightarrow \infty$, the expression $\frac{1}{n} \log_2 \frac{A_{\nd n}}{A_{\nd n -1}}$ converges to $\Gamma_n (\nd) - \Gamma_n (\nd -\frac{1}{n})$, being $\Gamma_n(\nd)$ given in \eqref{eq:proof_G_2}. From Lemma~\ref{corollary:der}  we know that  $\G'(\nd)> 0$ for $0 < \nd < 1/2$. As $n \rightarrow \infty$, from Theorem~\ref{theorem:growth_rate} we have $\Gamma_n(\nd) \rightarrow \G(\nd)$. Hence, for sufficiently large $n$, $\Gamma_n (\nd) \geq \Gamma_n (\nd -\frac{1}{n})$, and $A_{\d}$ is non decreasing.

We can now write
\begin{align}
\Pr\{ \D \leq \nd n\}  \leq  \nd n A_{\nd n} \leq\nd n 2^{n \Gamma_n(\nd) }\,  ,
\end{align}
where we have used $A_{\nd n}\leq 2^{n \Gamma_n(\nd) }$, being $\Gamma_n(\nd)$ given in \eqref{eq:proof_G_2}.

As $n \rightarrow \infty$ we have $\Gamma_n(\nd) \rightarrow \G(\nd)$. Moreover, $\G(\nd)<0$ for all $0 < \nd < \dmint$, provided $(\ri,\ro) \in \region$. Hence, $\Pr\{ \D \leq \nd n\}$ tends to $0$ exponentially on $n$.
\end{IEEEproof}
\begin{remark}
As from Theorem~\ref{theorem:min_dist}, we have an exponential decay of the probability to find codewords with weight less than $\dmint n$ when the $(\ri, \ro)$ pair belongs to the region $\region$. Such an exponential decay shall be attributed to the presence of the linear random outer code characterized by a dense parity-check matrix, which makes the growth rate function monotonically increasing for the values of $\nd$ for which it is negative.

As a comparison, for \ac{LDPC} code ensembles characterized by a positive normalized typical minimum distance, the growth rate function starts from $\G(0)=0$ with negative derivative, reaches a minimum, and then increases to cross the $x$-axis. In this case, for $\nd < \dmint$ the sum in the upper bound is dominated by those terms corresponding to small values of $w$, yielding either a polynomial decay (as for Gallager's codes \cite{Gallager63} ) or even  $\Pr \{ D \leq \nd n \}$ tending to a constant (as it is for irregular unstructured \ac{LDPC} ensembles \cite{orlitsky05:stopping,di06:weight}).
\end{remark}
\begin{theorem}\label{theorem:zero_codeword}
Let the random variable $\Z$ be the multiplicity of codewords of weight zero in the code book of a fixed-rate Raptor code picked randomly in the ensemble $\ensemble(\oensemble,\Omega, \ri, \ro, n)$. If $(\ri, \ro) \in \region$ then
\begin{align*}
\Pr \{ \Z > 1 \} \rightarrow 0 \quad \textrm{as } n\rightarrow \infty \, .
\end{align*}
\end{theorem}
\begin{IEEEproof}
In order to prove the statement we have to show that the probability measure of any event $\{ \Z = t \}$ with $t \in \mathbb N \setminus \{0, 1\}$ vanishes as $n \rightarrow \infty$. We start by analyzing the behavior of $\mathbb E[\Z]=A_0$, whose expression is $\mathbb E[\Z] = 1 + 2^{-n \ri (1 - \ro) } \sum_{l=1}^h \binom {h} {\l} (1 - \pl)^n$. Using an argument analogous to the one adopted in the proof of Theorem~~\ref{theorem:growth_rate}, for large enough $n$ we have
\begin{align*}
\frac{1}{n} \log_2 \left( 2^{-n \ri (1 - \ro) } \sum_{l=1}^h \binom {h} {\l} (1 - \pl)^n \right) \leq \Xi_n
\end{align*}
where
\begin{align*}
\Xi_n &:= - \ri (1 - \ro) + \frac{1}{n}\log_2(\ri n) + \sup_{\nl \in (0,1)} \Big\{ \ri \Hb(\nl)\\
& - \frac{1}{2n} \log_2 \left( 2 \pi \ri n \nl ( 1 - \nl) \right)
+ \log_2 (1 - \np_{\nl} + K/n) \Big\} \, .
\end{align*}
Therefore we can upper bound $\mathbb E[\Z]$ as $\mathbb E [\Z] \leq 1 + 2^{n \Xi_n}$ which, if $(\ri,\ro) \in \region$, implies $\mathbb E[\Z] \rightarrow 1$ exponentially as $n \rightarrow \infty$ due to $\Xi_n \rightarrow \G(0)$ and $\G(0)<0$.\footnote{It is worth noting that $\G(\nd)$ is right-continuous at $\nd=0$. This follows from the expression of $\G(\nd)$ proved in Theorem~\ref{theorem:growth_rate} and from the fact that $\fmax(\nd)$ is right-continuous at $\nd=0$ as shown in the proof of Theorem~\ref{theorem_inner}.} Next, it is easy to show that $\mathbb E[\Z] \geq 1$ and, via linear programming, that the minimum is attained if and only if $\Pr \{\Z=1\}=1$ and $\Pr\{\Z=t\}=0$ for all $t \in \mathbb N \setminus \{0,1\}$. Since in the limit as $n \rightarrow \infty$ of $\mathbb E[\Z] \rightarrow 1$, we necessarily have a vanishing probability measure for any event $\{ \Z = t \}$ with $t \in \mathbb N \setminus \{0, 1\}$.
\end{IEEEproof}
\begin{remark}
From Theorem~\ref{theorem:min_dist} and Theorem~\ref{theorem:zero_codeword}, a fixed-rate Raptor code picked randomly in the ensemble $\ensemble(\oensemble,\Omega, \ri, \ro, n)$ is characterized with probability approaching $1$ as $n \rightarrow \infty$ by a minimum distance at least equal to $\dmint n$ and by an encoding function whose kernel only includes the all-zero length $k$ message (hence bijective).
\end{remark}

In the following we introduce an outer region to $\region$ that only depends on the average output degree.
\begin{theorem}
\label{pro:outer}
The positive normalized typical minimum distance region $\region$ of a fixed-rate Raptor code ensemble $\msr{C}_{\infty}(\oensemble,\Omega, \ri, \ro)$ fulfills $\region \subseteq \outer$, where
\begin{equation}
\outer := \left\{(\ri,\ro) \succeq (0,0) | \ri \leq \min \left( \phi(\ro), \frac{1}{\ro}\right) \right\}
 \label{eq:outer_bound_2}
\end{equation}
with
\[
\phi(\ro)=
\begin{cases}
\frac{\bar \Omega \log_2 (1/\ro)}{\Hb(1-\ro) -(1-\ro)} \qquad 1 > \ro> \ro^* \\
1/\ro \qquad \qquad \qquad otherwise
\end{cases},
\]
being $\ro^*$ the only root of $\Hb(1-\ro) -(1-\ro)$ in  $\ro \in (0,1)$, numerically $\ro^* \approx 0.22709$.
\end{theorem}

\begin{IEEEproof}
See Appendix~\ref{sec:proof_outer}.
\end{IEEEproof}

%\begin{table}[th]
%\caption{Degree distributions $\Omega^{(1)}$ and $\Omega^{(2)}$}
%\begin{center}
%\begin{tabular}{|c|c|c|}
%\hline
%  Degree  &   $\Omega^{(1)}$ & $\Omega^{(2)}$ \rule{0pt}{2.6ex} \rule[-0.9ex]{0pt}{0pt} \\ \hline\hline
%  ${1}$ & 0.0098 & 0.0048 \\ \hline
%  ${2}$ & 0.4590 & 0.4965 \\ \hline
%  ${3}$ & 0.2110 & 0.1669 \\ \hline
%  ${4}$ & 0.1134 & 0.0734 \\ \hline
%  ${5}$ &  &  0.0822  \\ \hline
%  ${8}$ &  & 0.0575 \\ \hline
%  ${9}$ &  & 0.0360 \\ \hline
%  ${10}$ & 0.1113 &  \\ \hline
%  ${11}$ & 0.0799 &  \\ \hline
%  ${18}$ &  & 0.0012 \\ \hline
%  ${19}$ &  & 0.0543  \\ \hline
%  ${40}$ & 0.0156 &  \\ \hline
%  ${65}$ &  &  0.0182 \\ \hline
%  ${66}$ &  & 0.0091 \\ \hline \hline
%  $\bar \Omega$ & 4.6314 & 5.825  \rule{0pt}{2.6ex} \rule[-0.9ex]{0pt}{0pt} \\
%    \hline \hline
%\end{tabular}
%\end{center}\label{table:dist}
%\end{table}
\begin{table}[t]
\caption{Degree distributions $\Omega^{(1)}$, defined in \cite{MBMS12:raptor,luby2007rfc} and $\Omega^{(2)}$, defined in \cite{shokrollahi06:raptor}}
\begin{center}
\begin{tabular}{|c|c|c|c|c|c|c|c|c|c|c|c|c|c|c|c|}
%\begin{tabular}{|c|c|c|c|c|c|c|c|c|}
%\hline
%  Degree  &         $1$    & $2$    &   $3$&    $4$ & $5$   &  $8$  & $9$ & \\ \hline%& $10$   &   $11$& $18$ & $19$ & $40$   & $65$ & $66$ & $\bar \Omega$ \\ \hline \hline
%  $\Omega^{(1)}$ &  0.0098 & 0.459 & 0.211 & 0.1134 &       &      &      &\\ \hline%& 0.1113 & 0.0799&      &      & 0.0156 &      &      & 4.6314      \\ \hline
%  $\Omega^{(2)}$ &  0.0048 &0.4965 & 0.1669& 0.0734 &0.0822 &0.0575&0.036 & \\ \hline \hline%&        &       &0.0012&0.0543& 0.0156 &0.0182&0.0091& 5.825 \\ \hline
%  Degree         & $10$    & $11$  & $18$  & $19$   & $40$  & $65$ &  $66$ & $\bar \Omega$ \\ \hline
%  $\Omega^{(1)}$ & 0.1113  & 0.0799&       &        & 0.0156&      &       & 4.6314      \\ \hline
%  $\Omega^{(2)}$ &         &       &0.0012 &0.0543  & 0.0156&0.0182&0.0091 & 5.825 \\ \hline
\hline
  Degree& $\Omega^{(1)}$ & $\Omega^{(2)}$ \rule{0pt}{2.6ex} \rule[-0.9ex]{0pt}{0pt} \\ \hline\hline
  ${1}$ & 0.0098 & 0.0048 \\ \hline
  ${2}$ & 0.4590 & 0.4965 \\ \hline
  ${3}$ & 0.2110 & 0.1669 \\ \hline
  ${4}$ & 0.1134 & 0.0734 \\ \hline
  ${5}$ &  &  0.0822  \\ \hline
  ${8}$ &  & 0.0575 \\ \hline
  ${9}$ &  & 0.0360 \\ \hline
  ${10}$ & 0.1113 &  \\ \hline
  ${11}$ & 0.0799 &  \\ \hline
  ${18}$ &  & 0.0012 \\ \hline
  ${19}$ &  & 0.0543  \\ \hline
  ${40}$ & 0.0156 &  \\ \hline
  ${65}$ &  &  0.0182 \\ \hline
  ${66}$ &  & 0.0091 \\ \hline \hline
  $\bar \Omega$ & 4.6314 & 5.825  \rule{0pt}{2.6ex} \rule[-0.9ex]{0pt}{0pt} \\
    \hline
\end{tabular}
\end{center}\label{table:dist}
\end{table}

\begin{example}
\label{example_region}
In  Fig.~\ref{fig:region} we show the positive normalized typical minimum distance region, $\region$ for $\Omega^{(1)}$ and $\Omega^{(2)}$ (see Table~\ref{table:dist}) together with their outer bound $\outer$. It can be observed how the outer bound is tight in both cases except for inner codes rates close to $\ri=1$. The figure also shows several isorate curves, along which the rate of the Raptor code $\rate=\ri~\ro$ is constant. For example, in order to have a positive normalized typical minimum distance and an overall rate $\rate=0.95$, the figure shows that the rate of the ouer code must lay below $\ro<0.978$ for both distributions. Let us assume we want to design a fixed-rate Raptor code, with degree distribution $\Omega^{(1)}$ or $\Omega^{(2)}$, overall rate $\rate=0.95$ and for a given length $n$, which we assume to be large. Different choices for $\ri$ and $\ro$ are possible. If $\ro$ is not chosen as $\ro<0.978$ the average minimum distance of the ensemble will not grow linearly on $n$. Hence, many codes in the ensemble will exhibit high error floors even under \ac{ML} erasure decoding.
\begin{figure}[h!]
        \centering
         \begin{subfigure}[b]{0.48\textwidth}
        \includegraphics[width=1.05\textwidth]{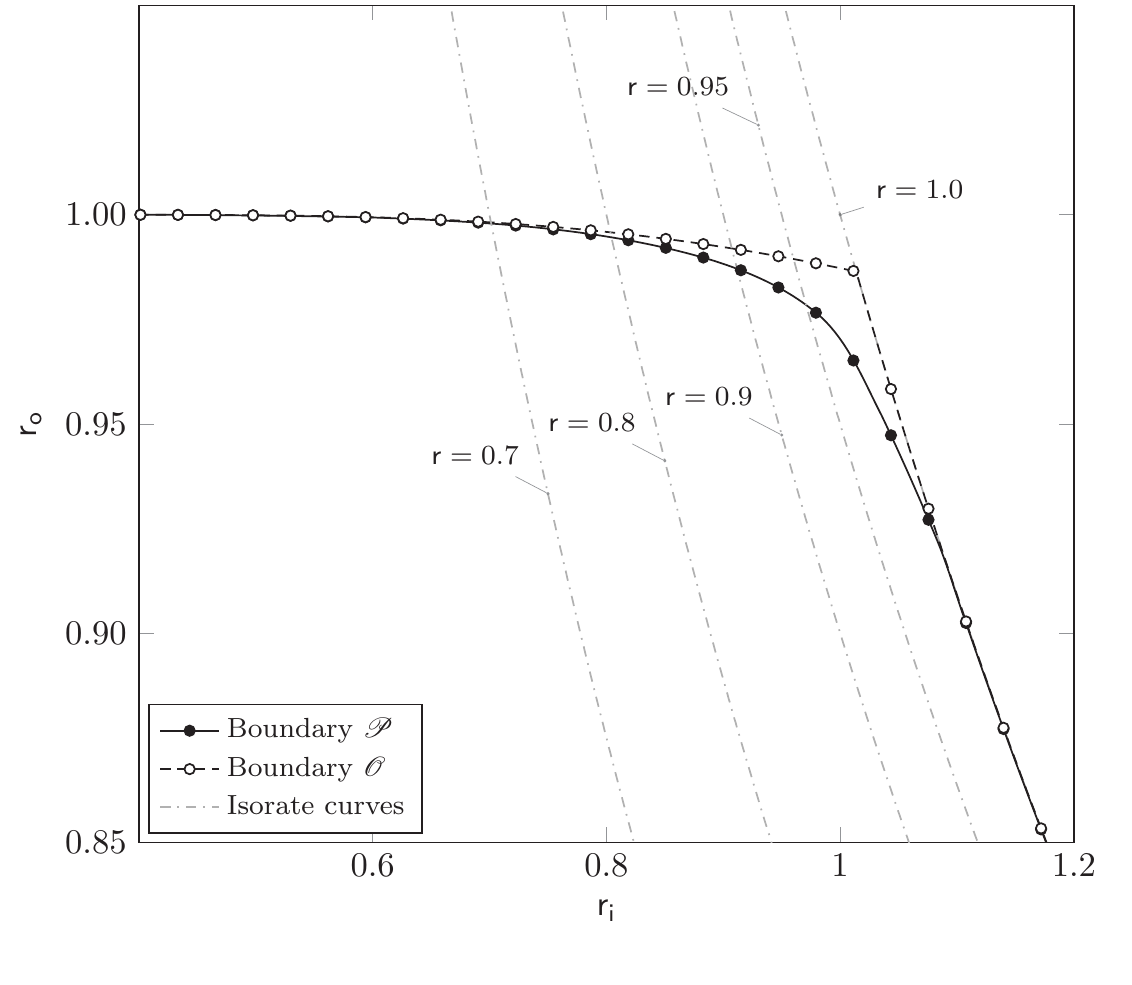}
        \subcaption{$\Omega^{(1)}$}
        \label{fig:region_a}
        \end{subfigure}
        \begin{subfigure}[b]{0.48\textwidth}
        \includegraphics[width=1.05\textwidth]{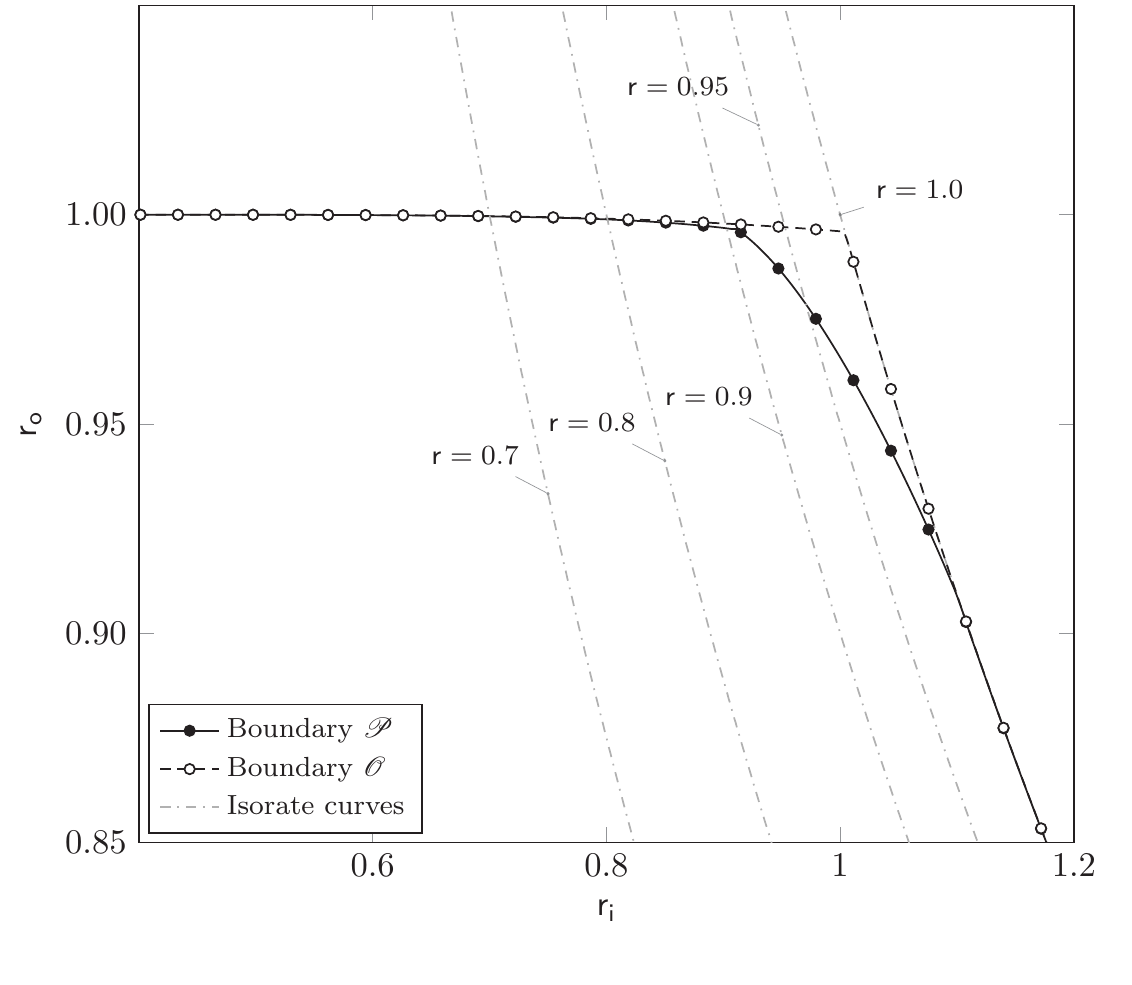}
        \subcaption{$\Omega^{(2)}$}
        \label{fig:region_b}
        \end{subfigure}
        \caption{Positive growth rate region. The solid lines with black markers represent the positive growth-rate $\region$ and the dashed lines with white markers represents its outer bound $\outer$. The gray dashed lines represent isorate curves for different rates $\rate$.}
\label{fig:region}
\end{figure}
\end{example}

\section{Finite-Length Results}\label{sec:results}

In this section experimental results are presented to validate the analytical results obtained in the previous sections. By means of examples we illustrate how the developed results can be used to make accurate statements about the performance of fixed-rate Raptor code ensembles in the finite length regime.
Furthermore, we provide some results that show a tradeoff between performance and decoding complexity.
%Furthermore we will show how in order to design good Raptor code ensembles for finite length it is not enough to look only at the growth rate. %One should rather look at the typical minimum distance of the ensemble considering also the complexity of decoding, which under inactivation decoding depends strongly on the number of inactivations.
Finally we present some simulation results that show that the results obtained for linear random outer codes are a fair approximation for the results obtained with the standard R10 Raptor outer code (see \cite{MBMS12:raptor,luby2007rfc}).

\subsection{Results for Linear Random Outer Codes}\label{sec:results_CER}

\begin{figure}[t]
        \centering
        \begin{subfigure}[b]{0.49\textwidth}
            \includegraphics[width=0.98\columnwidth]{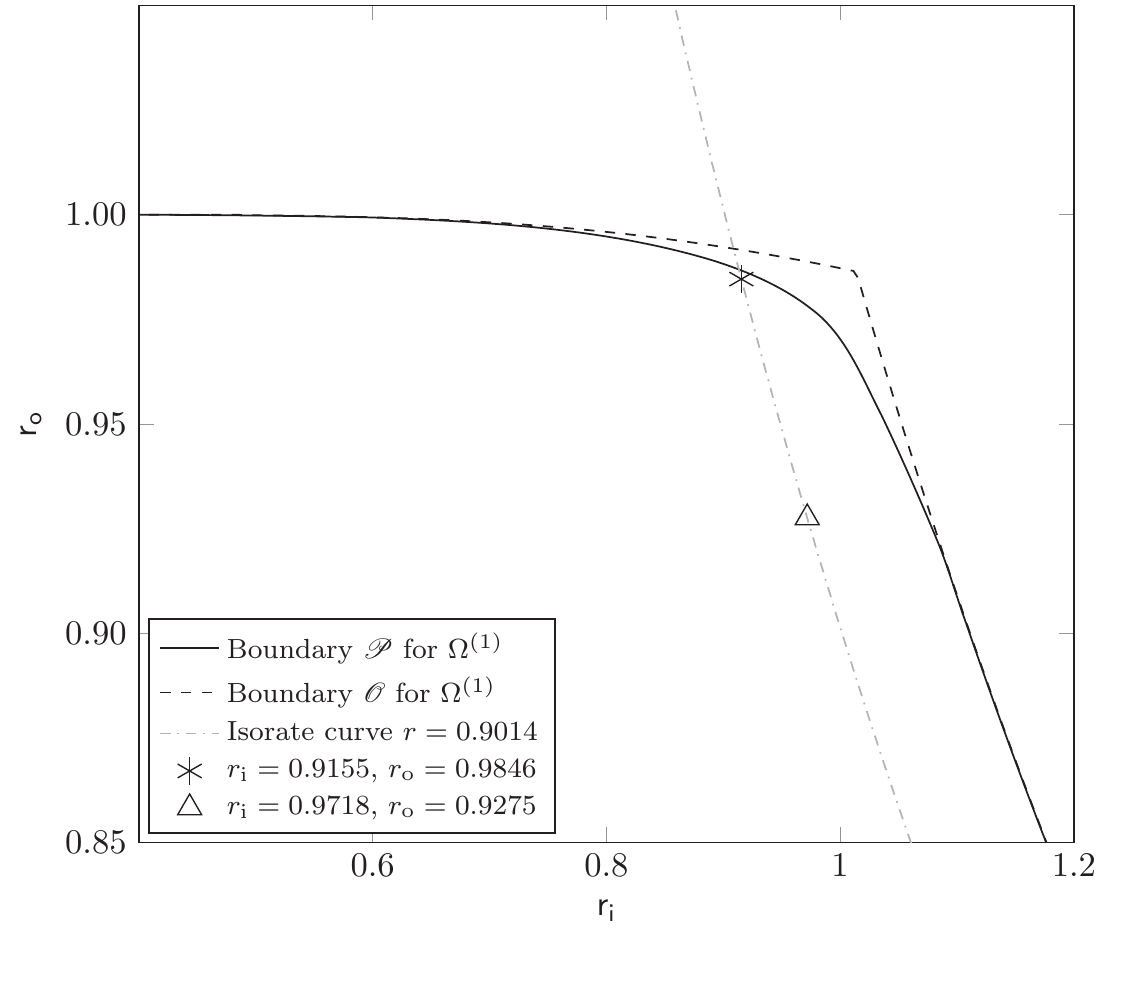}
            \subcaption{growth rate region}\label{fig:region_results}
        \end{subfigure}
        \begin{subfigure}[b]{0.49\textwidth}
            \raisebox{1.8mm}{\includegraphics[width=0.98\columnwidth]{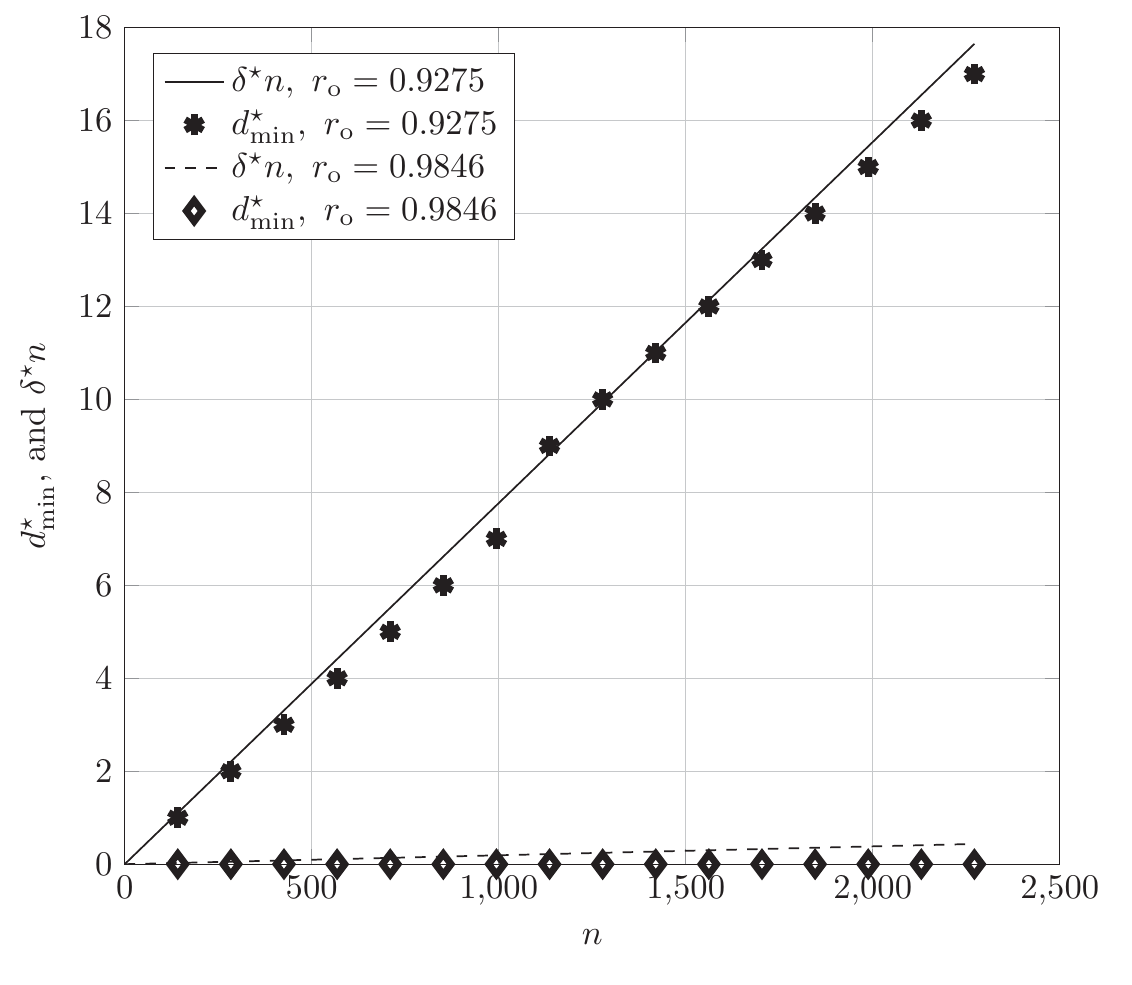}}
            \subcaption{minimum distance}\label{fig:d_min_delta_star}
        \end{subfigure}
        \caption{In the upper figure the solid and dashed lines represent the positive growth rate region of $\Omega^{(1)}$ its outer bound. The dashed-dotted line represents the isorate curve for $\rate=0.9014$ and the markers represent two different points along the isorate curve
         with the same rate $\rate$ but different values of $\ri$ and $\ro$.
        The lower figure shows the typical minimum distance $\dmintt$ as a function of the blocklength $n$ for ensembles with $\ro= 0.9275$ and $\ro=0.9846$ and $\rate=0.9014$. The markers represent $\dmintt$ whereas the lines represent $\dmint n$.}\label{fig:sims}
\end{figure}

In this section we will consider Raptor code ensembles $\ensemble(\oensemble,\Omega^{(1)}, \ri, \ro, n)$ for different values of $\ri$, $\ro$, and $n$ but keeping the overall rate of the Raptor code constant to $\rate=0.9014$. Fig.~\ref{fig:region_results} shows the boundary of $\region$ and $\outer$ for \ac{LT} distribution $\Omega^{(1)}$ together with an isorate curve for $\rate=0.9014$. The markers along the isorate curve in the figure represent the two different combinations of $\ri$ and $\ro$ that will be considered in this section. The first point ($\ri=0.9155$, $\ro=0.9846$), marked with an asterisk, is inside but very close to the boundary of $\region$ for $\Omega^{(1)}$. We will refer to ensembles corresponding to this point as \emph{bad} ensembles. The second point, ($\ri=0.9718$, $\ro=0.9275$) marked with a triangle, is inside and quite far from the boundary of $\region$ for $\Omega^{(1)}$. We will refer to ensembles corresponding to this point as \emph{good} ensembles.

In order to analyze ensembles of finite length Raptor codes it is useful to introduce a notion of minimum distance for finite length.
\begin{mydef}
The typical minimum distance, $\dmintt$  of an ensemble $\ensemble(\oensemble,\Omega, \ri, \ro, n)$ is defined as the integer number
\begin{align*}
\dmintt := \begin{cases}
0 & \mkern-48mu \text{if } A_0 > 1 + 1/2 \\
\max \{ d\geq0 : \left(\sum_{i=0}^{d} \we_i -1\right) < 1/2 \} & \text{otherwise.}
\end{cases}
\end{align*}
\end{mydef}
This definition will come in handy when we expurgate Raptor code ensembles. In fact, at least half of the codes in the ensemble will have a minimum distance of $\dmintt$ or larger. The equivalent of $\dmintt$ in the asymptotic regime is $\dmint$, the (\emph{asymptotic}) normalized minimum distance of the ensemble $\msr{C}_{\infty}(\oensemble,\Omega, \ri, \ro)$.  For sufficiently large $n$ one expects that $\dmintt$ converges to $\dmint n$. Fig.~\ref{fig:d_min_delta_star} shows $\dmintt$ and $\dmint n$ as a function of the blocklength $n$. It can be observed how the good ensemble has a larger typical minimum distance than the bad ensemble. In fact for all values of $n$ shown in Fig.~\ref{fig:d_min_delta_star} $\dmintt=0$ for the bad ensemble. We can also see how already for small values of $n$ the $\dmintt$ and $\dmint n$ are very similar. Hence, the result of our asymptotic analysis of the minimum distance holds already for small values of $n$.

%\subsection{CER Results}\label{sec:results_CER}
The expression of the average weight enumerator in Theorem~\ref{theorem:we} can be used in order to upper bound the average \acf{CER}  over a \ac{BEC} with erasure probability $\epsilon$, \cite{CDi2001:Finite}. However, the upper bound proposed in \cite{CDi2001:Finite} needs to be slightly modified to take into account codewords of weight $0$. We have

%\begin{equation}
\begin{align}
\label{eq:bound_Gavg}
%\begin{array}{ll}
&\Bbb{E}_{\ensemble(\oensemble,\Omega, \ri, \ro, n)} \left[P_B(\epsilon)\right]\leq
P^{(\mathsf S)}_{B}(n,k,\epsilon) \nonumber \\
& + \sum_{e=1}^{n-k} {n \choose e} \epsilon^e (1-\epsilon)^{n-e} \min \left\{1, \sum_{w=1}^e {e \choose w} \frac{\we_w}{{n \choose w}}\right\} + \we_0-1
%\end{array}
%\end{equation}
\end{align}
where $P^{(\mathsf S)}_{B}(n,k,\epsilon)$ is the Singleton bound
\begin{equation}\label{eq:bound_S}
P^{(\mathsf S)}_{B}(n,k,\epsilon)= \sum_{e=n-k+1}^n {n \choose e} \epsilon^e (1-\epsilon)^{n-e}.
\end{equation}

Considering Raptor codes in a fixed-rate setting also allows us to expurgate Raptor code ensembles as it was done in \cite{Gallager63} for LDPC code ensembles. Let us consider an integer
$\ds \geq 0$ so that
\begin{align}
\label{eq:pr_d_min_ex}
\Pr\{ \dmin \leq \ds\} & \leq   \sum_{w=0}^{\ds} A_w -1 = \theta < 1/2.
\end{align}
We can define the expurgated ensemble $\ensemble^{\text{ex}}(\oensemble,\Omega, \ri, \ro, n, \ds)$ as the ensemble of codes in the ensemble $\ensemble(\oensemble,\Omega, \ri, \ro, n)$  whose minimum distance is $\dmin >\ds$. The expurgated ensemble will contain a fraction at least  $1 - \theta>1/2$ of the codes in the original ensemble. From \cite{Gallager63} it is known that the average \ac{WE} of the expurgated ensemble can be upper bounded by:
\begin{align*}
\we^{\text{ex}}_d
\begin{cases}
\leq  2 \we_d  & \text{if } d > \ds  \\
= 0             & \text{if } 1 \leq d \leq\ds \\
\end{cases}
\end{align*}

For each ensemble considered in this section  $6000$ codes\footnote{For clarity of presentation only 300 codes are shown in the figures.} were selected randomly from the ensemble. For each code Monte Carlo simulations over a \ac{BEC} were performed  until $40$ errors were collected or a maximum of $10^5$ codewords were simulated. We remark that the objective here was not so much characterizing the performance of every single code but rather to characterize the average performance of the ensemble.

\begin{figure}[t]
        \centering
        \begin{subfigure}[b]{0.49\textwidth}
            \includegraphics[width=0.99\columnwidth]{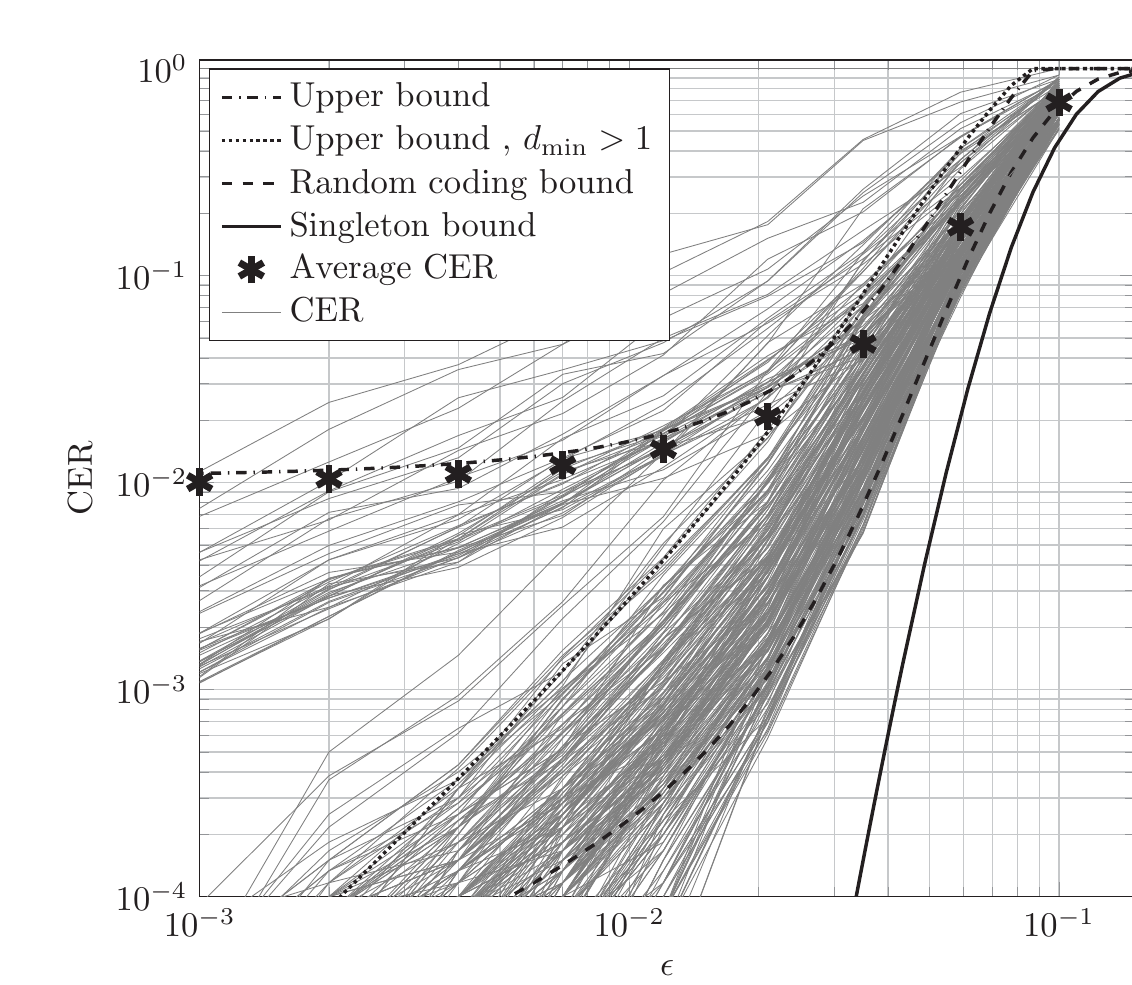}
            \subcaption{$\ro=0.9275$, $\rate=0.9014$} \label{fig:UB_good_128}
        \end{subfigure}
        \begin{subfigure}[b]{0.49\textwidth}
            \includegraphics[width=0.99\columnwidth]{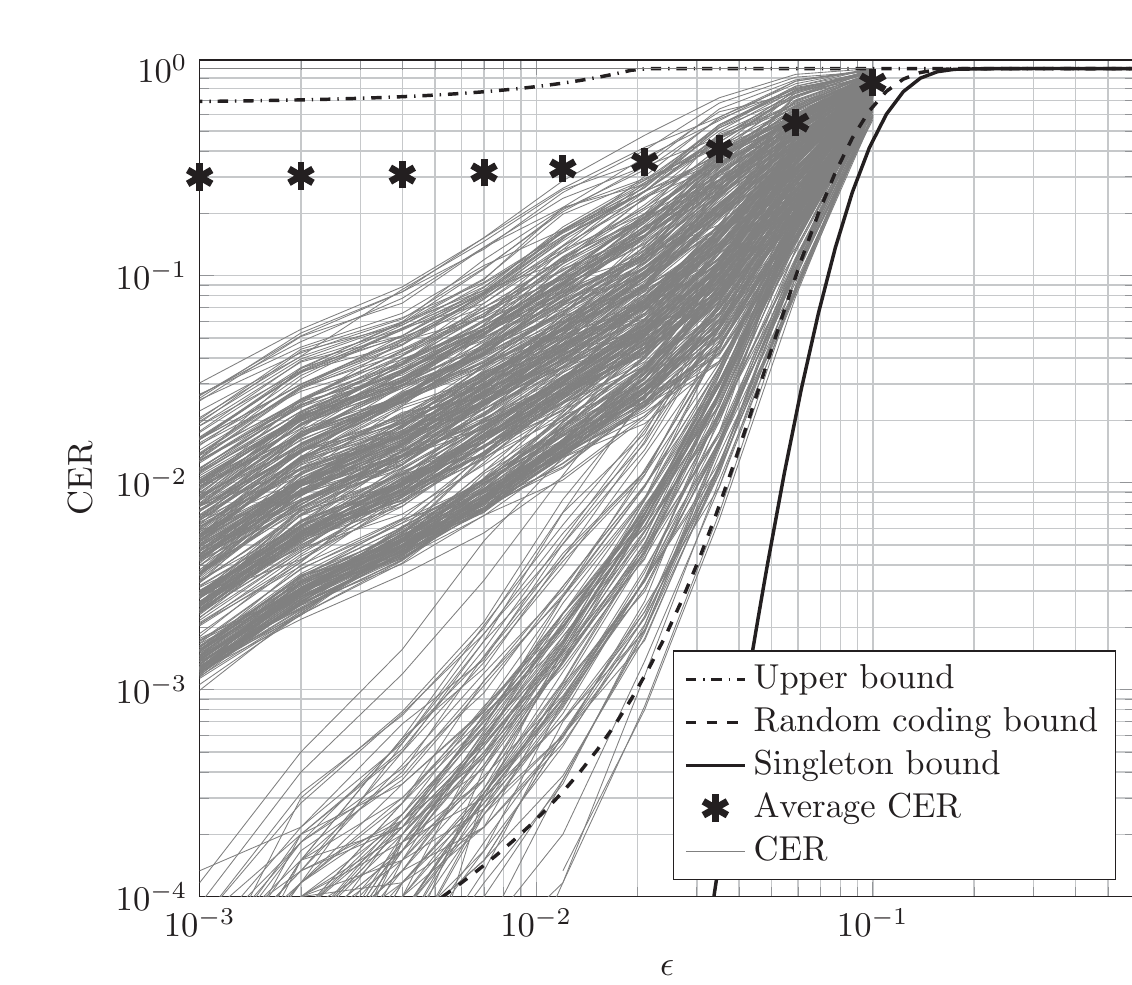}
            \subcaption{$\ro=0.9846$, $\rate=0.9014$}\label{fig:UB_bad_128}
        \end{subfigure}
        \caption{Codeword error rate \ac{CER} vs erasure probability $\epsilon$ for two ensembles with $\rate=0.9014$ and $k=128$ but different values of $\ro$.  The solid, dashed and dot-dashed lines represent respectively the Singleton bound, the Berlekamp random coding bound and the upper bound in. The dotted line represents the upper bound for the expurgated ensemble for $\ds=1$. The markers represent the average \ac{CER} of the ensemble and the thin gray curves represent the performance of the different codes in the ensemble, both obtained through Monte Carlo simulations.}\label{fig:UB_128}
\end{figure}

Fig.~\ref{fig:UB_128} shows the \ac{CER} vs the erasure probability $\epsilon$ for two ensembles with $\rate=0.9014$ and $k=128$ that have different outer code rates, $\ro=0.9275$ (good ensemble) and $\ro=0.9846$ (bad ensemble). The good ensemble is characterized by a typical minimum distance $\dmintt=2$ whereas the bad ensemble is characterized by $\dmintt=0$ (cf. Fig.~\ref{fig:d_min_delta_star}).
For the two ensembles the upper bound in \eqref{eq:bound_Gavg} holds for average CER. However, the performance of the codes in the ensemble shows a high dispersion due to the short blocklength ($n=142$). In fact in both ensembles there are codes with minimum distance equal to zero which have \ac{CER}$=1$ (around $1\%$ for the good ensemble and $30\%$ for the bad ensemble).
%It can be observed how the upper bound in \eqref{eq:bound_Gavg} holds in the two cases. In both cases there are codes in the ensemble with minimum distance equal to zero which have \ac{CER}$=1$ for all erasure rates. In the good ensemble approximately $1\%$ of the codes have zero minimum distance whereas in the bad ensemble it is $30\%$. The performance of the codes in the ensemble shows a high dispersion due to the short blocklength ($n=142$).
Comparing Fig.~\ref{fig:UB_good_128} and Fig.~\ref{fig:UB_bad_128} one can easily see how  the fraction of codes performing close to the random coding bound is larger in the good ensemble than in the bad ensemble. For the good ensemble Fig.~\ref{fig:UB_good_128} shows also an upper bound on the average \ac{CER} for the expurgated ensemble with $\ds=1$, that has a lower error floor.
%It can be observed how the error floor can be substantially decreased  by expurgation.
%For the bad ensemble there exists no $\ds\geq 0$ that leads to $\theta<1/2$ in \eqref{eq:pr_d_min_ex}, and hence no expurgated ensemble can be defined following the approach in \cite{Gallager63}.
For the bad ensemble no expurgated ensemble can be defined (no $\ds\geq 0$ exists that leads to $\theta<1/2$ in \eqref{eq:pr_d_min_ex}).

\begin{figure}[t]

        \centering
        \begin{subfigure}[b]{0.49\textwidth}
            \includegraphics[width=0.99\columnwidth]{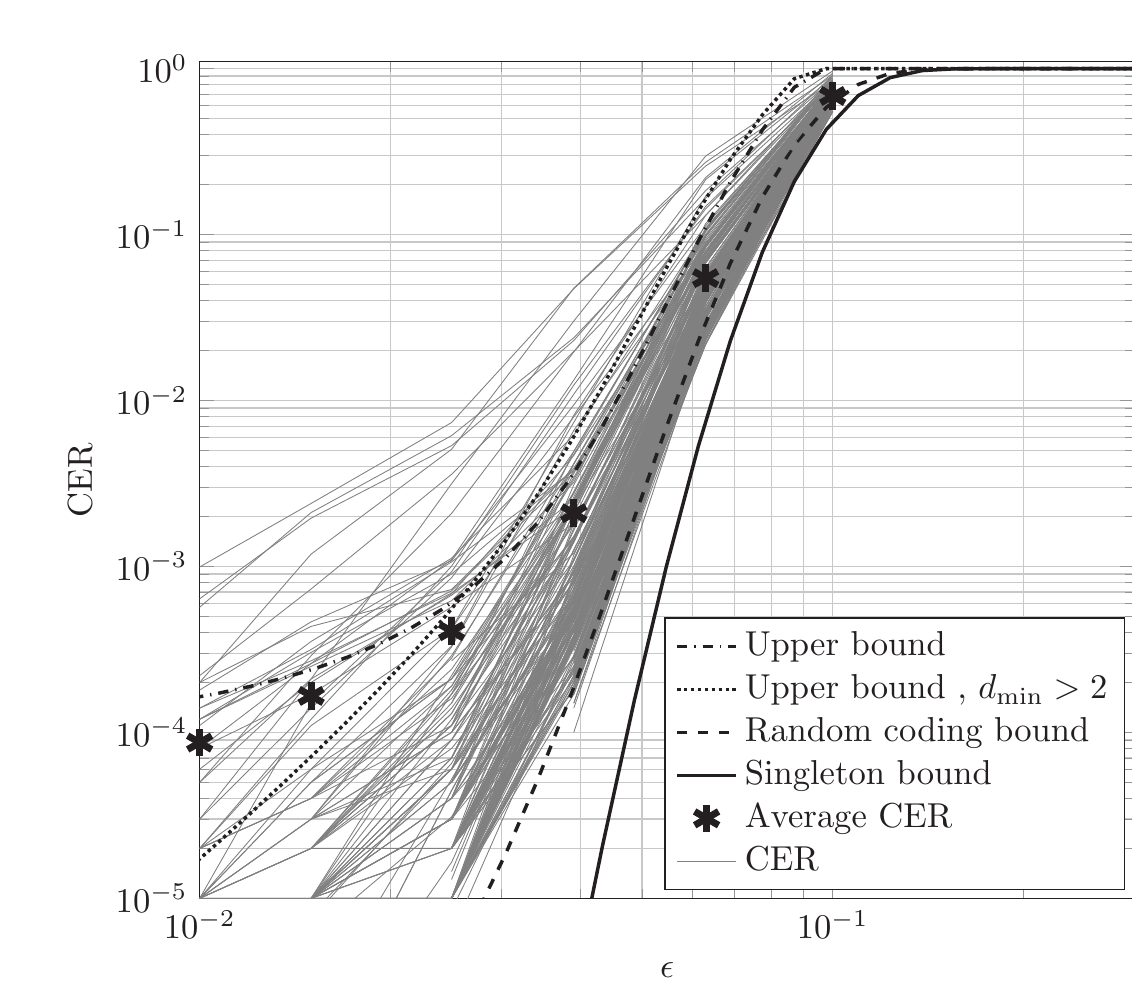}
            \subcaption{$\ro=0.9275$, $\rate=0.9014$} \label{fig:UB_good_256}
        \end{subfigure}
        \begin{subfigure}[b]{0.49\textwidth}
            \includegraphics[width=0.99\columnwidth]{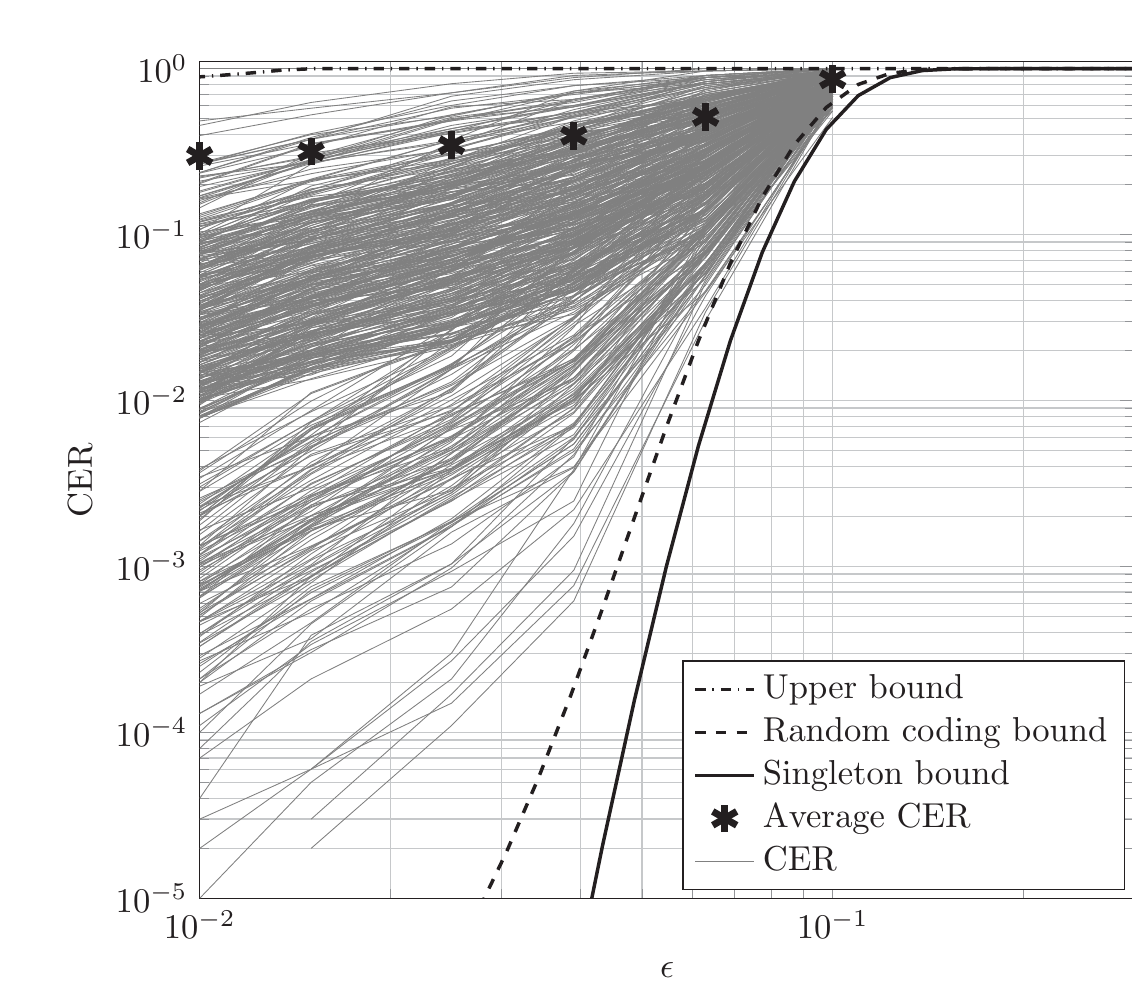}
            \subcaption{$\ro=0.9846$, $\rate=0.9014$}\label{fig:UB_bad_256}
        \end{subfigure}
        \caption{Codeword error rate \ac{CER} vs erasure probability $\epsilon$ for two ensembles with $\rate=0.9014$ and $k=256$ but different values of $\ro$.  The solid, dashed and dot-dashed lines represent respectively the Singleton bound, the Berlekamp random coding bound and the upper bound. The dotted line represents the upper bound for the expurgated ensemble for $\ds=2$. The markers represent the average \ac{CER} of the ensemble and the thin gray curves represent the performance of the different codes in the ensemble, both obtained through Monte Carlo simulations.}\label{fig:UB_256}

\end{figure}

Fig.~\ref{fig:UB_256} shows the \ac{CER} vs $\epsilon$ for two ensembles using the same outer code rates as in Fig.~\ref{fig:UB_128} but this time for $k=256$. It can be observed how the \ac{CER} shows somewhat less dispersion than for $k=128$. If we compare Fig.~\ref{fig:UB_good_256} and Fig.~\ref{fig:UB_good_128} we can see how for the good ensemble ($\ro=0.9275$) the error floor is much lower for $k=256$ than for $k=128$, due to an increase in the typical minimum distance. In fact, whereas for $k=128$ there were some codes with minimum distance zero for $256$ we did not find any code with minimum distance zero out of the $6000$ codes which were simulated. For the good ensemble it is possible again to considerably lower the error floor by expurgation.
However, comparing Fig.~\ref{fig:UB_bad_256} and Fig.~\ref{fig:UB_bad_128} we can see how the error floor is approximately the same for $k=128$ and $k=256$, because in both cases  the typical minimum distance is zero.
%Comparing Fig.~\ref{fig:UB_bad_256} and Fig.~\ref{fig:UB_bad_128} we can see how the error floor is approximately the same for $k=128$ and $k=256$. In fact, for both cases the typical minimum distance is zero. For the bad ensemble, also for $k=256$ it is not possible to define an expurgated ensemble as introduced in \cite{Gallager63}.

\begin{figure}[t]
        \centering
        \includegraphics[width=\figwidth]{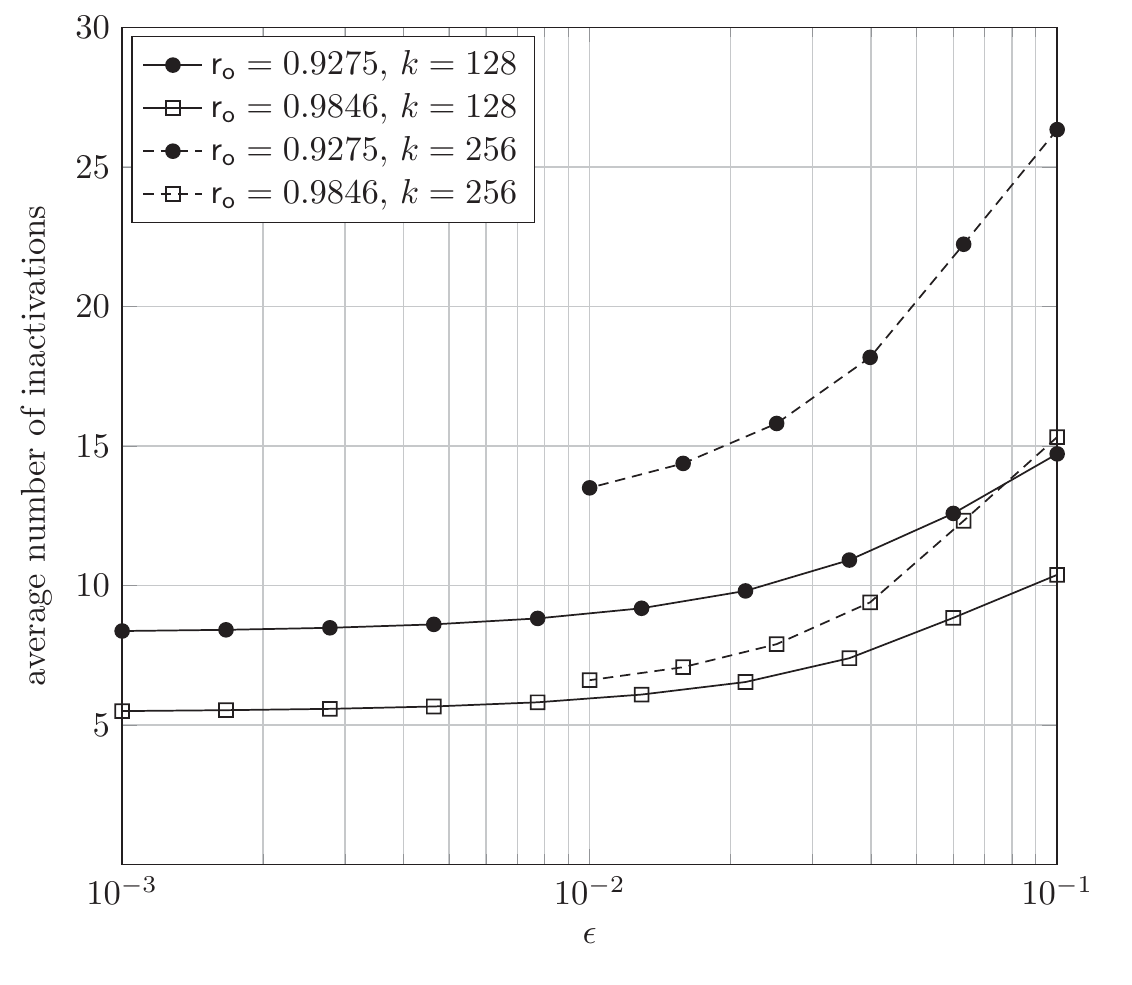}
        \caption{Average number of inactivations vs. erasure probability of the channel. The dot markers stand for an outer code rate $\ro=0.9275$ and the square markers for $\ro=0.9846$. The solid line stands for $k=128$ and the dashed line for $k=256$.}
\label{fig:inact}
\end{figure}

%\subsection{Decoding Complexity}
So far we have only considered the \ac{CER} performance under \ac{ML} decoding. In practical systems one needs to consider decoding complexity as well.
When inactivation decoding is used the decoding throughput is largely determined by the number of inactivations needed for decoding \cite{shokrollahi2005systems}, since the decoding complexity  is cubic in the number of inactivations. Fig.~\ref{fig:inact} shows the averaged number of inactivations needed for ensembles of Raptor codes with output degree distribution  section.
%The results were obtained by means of simulations averaging over the $6000$ codes simulated in Section \ref{sec:results_CER}.
It can be observed how the good ensembles ($\ro=0.9275$) need more inactivations than bad ensembles ($\ro=0.9846$). Hence, the better \ac{CER} performance obtained by using an outer code with lower rate comes at the cost of a higher decoding complexity.

\subsection{Comparison with Raptor Codes with a Standard R10 Outer Code}
\begin{figure}[t]
        \centering
        \begin{subfigure}[b]{0.49\textwidth}
            \includegraphics[width=0.99\columnwidth]{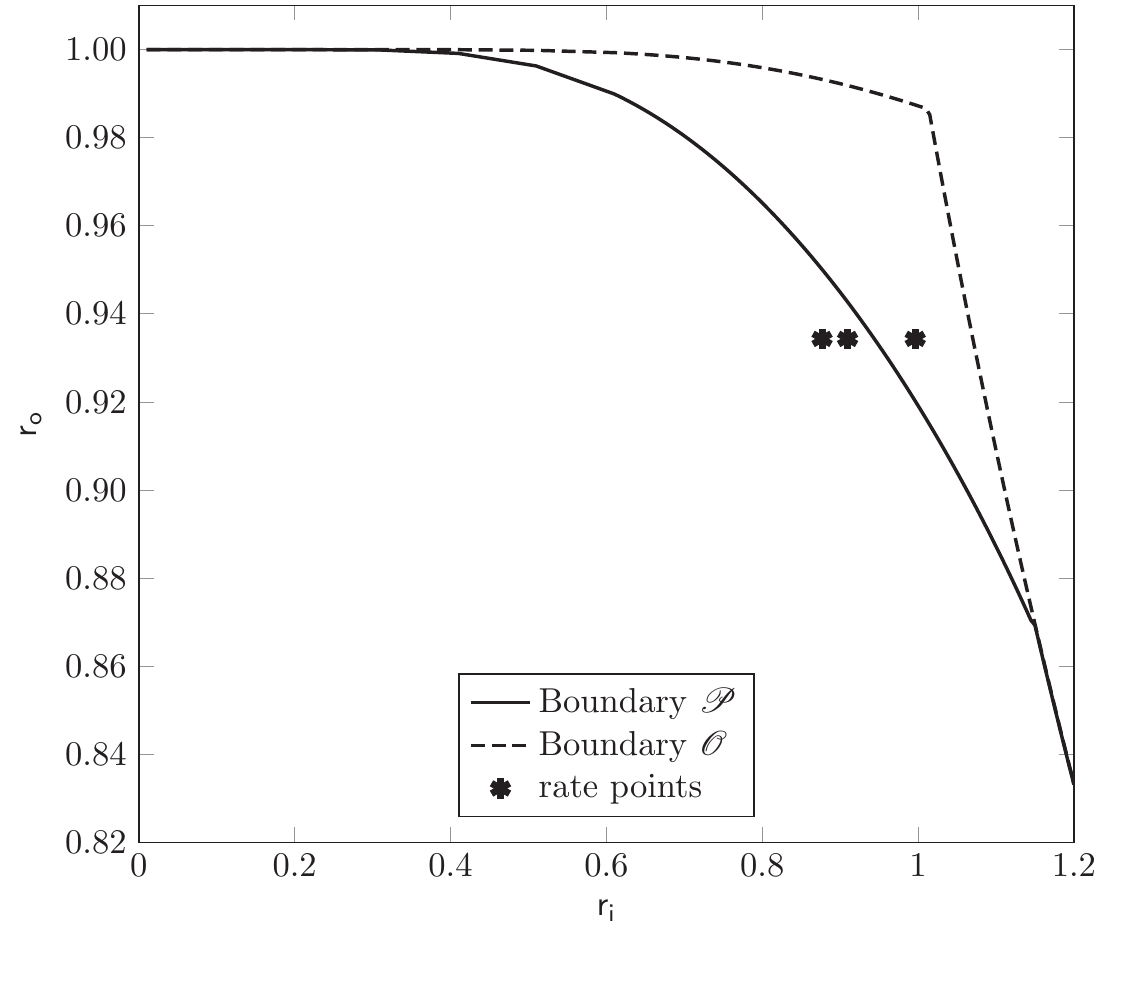}
            \subcaption{positive growth rate region} \label{fig:region_compare}
        \end{subfigure}
        \begin{subfigure}[b]{0.49\textwidth}
            \raisebox{1.1mm}{\includegraphics[width=0.99\columnwidth]{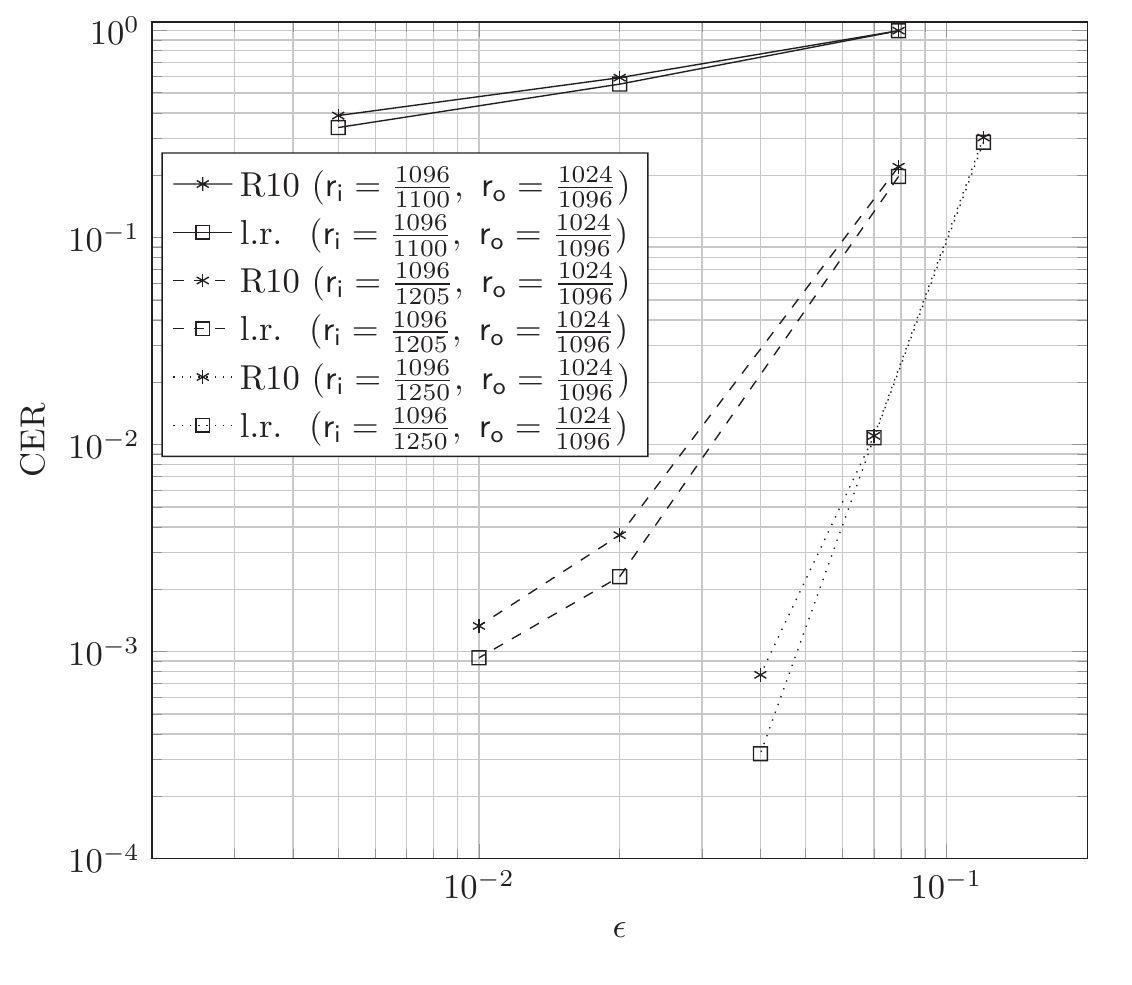}}
            \subcaption{CER results}\label{fig:CER_compare}
        \end{subfigure}
        \caption{The upper figure shows the positive growth rate region for the degree distribution  $\Omega(x) = 0.0098 x   + 0.4590  x^2 +  0.2110  x^3 +  0.1134 x^4 +   0.2068 x^5$.  The markers represent three different rate points all of them with $\ro=1024/1096$ but with different inner code rates,    $\ri=1096/1100$, $\ri=1096/1205$ and $\ri=1096/1250$. The lower figure shows the average \ac{CER} for Raptor code ensembles using $\Omega(x)$ as output degree distribution and two different outer codes, the standard outer code of R10 Raptor codes and a linear random outer code, (l.r.) in the legend.}\label{fig:r10_compare}
\end{figure}

In this section we illustrate by means of a numerical example how the results obtained for linear random outer code closely approximate the results with the standard R10 Raptor outer code  \cite{MBMS12:raptor}\cite{luby2007rfc}. We consider Raptor codes with an \ac{LT} degree distribution $\Omega(x) = 0.0098 x   + 0.4590  x^2 +  0.2110  x^3 +  0.1134 x^4 +   0.2068 x^5$. Fig.~\ref{fig:region_compare} shows the positive growth rate region for such a degree distribution (assuming a linear random outer code) and the three different rate points, two of which are inside the region $\region$ while the third one lays outside. The $(\ri,\ro)$ rate pairs for the three points are specified in the figure caption.

Fig.~\ref{fig:CER_compare} shows the average \ac{CER} obtained through Monte Carlo simulations for the ensembles of Raptor codes with $k=1024$, output degree distribution $\Omega(x)$ and two different outer codes, the standard R10 outer code and a linear random outer code. For the three rate points considered the average CER using the standard outer code and a linear random outer code are very close.
As it can be observed, the error floor behavior of the Raptor code ensemble with R10 outer code is in agreement with the  position of the corresponding point on the $(\ri,\ro)$ plane with respect to the $\region$ region, although this region is obtained using the simple linear random outer code model. For rate points inside $\region$ the error floor is much lower, and it tends to become lower the further the point is from the boundary of $\region$.
%Furthermore, as it can be observed in Fig.~\ref{fig:CER_compare} for a rate point outside the positive growth rate region $\region$ the \ac{CER} is very high. However for rate points inside $\region$ the \ac{CER} is much lower, and it becomes lower the further the rate point is from the boundary of $\region$.

\section{Conclusions}\label{sec:Conclusions}
In this work we have considered ensembles of binary fixed-rate Raptor codes which use linear random codes as outer codes. We have derived the expression of the average \acl{WE} of an ensemble and the expression of the growth rate of the \acl{WE} as functions of the rate of the outer code and the rate and degree distribution of the inner \ac{LT} code.
Based on these expressions we are able to determine necessary and sufficient conditions to have Raptor code ensembles with a positive typical minimum distance. A simple necessary condition has been developed too, which only requires  (besides the inner and outer code rates) the knowledge of the average output degree. Simulation results have been presented that demonstrate the applicability of the theoretical results obtained for finite length Raptor codes. Moreover, simulation results have been presented that show that the performance of Raptor codes with  linear random outer codes is close to that of Raptor codes with the standard outer code of R10 Raptor codes. Thus, we speculate that the results obtained for Raptor codes with linear random outer codes hold as first approximation for standard R10 Raptor codes.

The work presented in this paper helps to understand the behavior of fixed-rate Raptor codes under \ac{ML} decoding and can be used to design Raptor codes for \ac{ML} decoding. Despite the fact that only the fixed-rate setting has been considered, we speculate that Raptor code ensembles with a good fixed-rate performance will have also a good performance in a rateless setting.

Although only binary Raptor codes have been considered, the authors believe that the work can be extended to non-binary Raptor codes with a limited effort.

\appendices
\section{Proof of Theorem \ref{theorem_inner}}\label{sec:proof_inversion}

We will first prove that for all $(\ri,\ro)$ pairs in $\region$ we have a
 positive normalized typical minimum distance. Then we will prove that this is not possible for any other $(\ri,\ro)$ pair.

\subsection{Proof of Sufficiency}
A sufficient condition for a positive normalized typical minimum distance is
\begin{equation}
\lim_{\nd \to 0^+} \G(\nd) < 0
\end{equation}
which, from Theorem \ref{theorem:growth_rate}, is equivalent to
\begin{equation}
\ri  (1-\ro) >  \lim_{\nd \to 0^+} \max_{ \nl \in \mathscr D_{\nl}} \f(\nd, \nl).
\label{eq:lim_G}
\end{equation}
As done in Lemma~\ref{lemma:growth_rate_derivative} and Lemma~\ref{corollary:der}, let us  use the notation $\np(\nl)= \npnl$ to emphasize the dependence on $\nl$. We now show that
\begin{align} \label{eq:lim_max}
\lim_{\nd \to 0^+} \max_{ \nl \in \mathscr D_{\nl}} \f(\nd, \nl) &= \max_{ \nl \in \mathscr D_{\nl}} \lim_{\nd \to 0^+} \f(\nd, \nl) %\\
  = \max_{ \nl \in \mathscr D_{\nl}}  \left[ \ri \Hb(\nl) + \log_2 \left(1 - \np(\nl) \right)  \right]
\end{align}
that is we can invert maximization with respect to $\lambda$ and limit as $\delta \rightarrow 0^+$,  so that the region $\region$ in \eqref{eq:theorem_region} is obtained.
%\end{lemma}

This fact is proved by simply showing that
\begin{align}
\lim_{\nd \rightarrow 0^+} \fmax(\nd) =  \fmax(0),
\end{align}
that is the function $\fmax(\nd)=\max_{ \nl \in \mathscr D_{\nl}} \f(\nd,\nl)$ is right-continuous at $\nd = 0$. It suffices to show
\begin{align}\label{eq:max_a_b}
\fmax(\nd) = \max_{\nl \in (a, b)} \f(\nd, \nl)
\end{align}
where $(a,b)$ is an interval independent of $\nd \in [0,\frac{1}{2})$ and such that the function
\begin{align}
\log_2 \np(\nl)-\log_2(1-\np(\nl))
\end{align}
 is bounded over it, i.e.,
$$
\sup_{\nl \in (a,b)} \left| \log_2 \npnl - \log_2(1-\npnl) \right| = K \, .
$$
In fact, under these conditions we have uniform convergence of $\f(\nd,\nl)$ to $\f(0,\nl)$ in the interval $(a,b)$ as $\nd \rightarrow 0^+$, namely,
\begin{align}\label{eq:uniform_convergence}
\f(0,\nl) - K \nd \leq \f(\nd,\nl) \leq \f(0,\nl) + K \nd, \\
\quad \quad \quad \forall \lambda \textrm{ s.t. } a < \nl < b \,
\,
\end{align}
The second inequality in \eqref{eq:uniform_convergence} implies $\fmax(\nd) \leq \fmax(0) + K \nd$. Moreover, denoting by $\hat{\nl} \in (a,b)$ the maximizing $\nl$, we have
$$
\fmax(0) - K \nd = \f(0,\hat{\nl}) - K \nd \leq \f(\nd,\hat{\nl})
$$
which implies $\fmax(0) - K \nd \leq \fmax(\nd)$. So we have
$$
\fmax(0) - K \nd \leq \fmax(\nd) \leq \fmax(0) + K \nd
$$
which yields $\lim_{\nd \rightarrow 0^+} \fmax(\nd) = \fmax(0)$ as desired.

Next, we prove \eqref{eq:max_a_b}. We first observe that %
%$\max_{\nl \in (0,1)} \pnl = 1$ if and only if
in the case $\Omega_j=0$ for all even $j$ (in which case $\np(\nl)$ is strictly increasing) by direct computation we have $\partial\, \f(\nd,\nl) / \partial \nl < 0$ for all $0 \leq \nd < 1/2$ and for all $1/2 \leq \nl < 1$. Hence in this case we can take $b=1/2$. In all of the other cases there exists $\xi$ such that $\np(\nl) \leq \xi < 1$ for all $0 < \nl < 1$ and we can take $b=1$. The existence of $0 < a < 1/2$ (independent of $0 \leq \nd < 1/2$) such that the maximum is not taken for all $0 < \nl \leq a$ is proved as follows. Denoting $c = \log_2 e$ and $\np'(\nl)=\mathrm d \np(\nl) / \mathrm d \nl$, we have
\begin{align*}
\frac{\partial\, \f(\nd,\nl)}{\partial \nl} = \ri \log_2(1-\nl) - \ri \log_2 \nl \\
+ c\, \nd \, \frac{\np'(\nl)}{\np(\nl)} - c\,(1-\nd) \frac{\np'(\nl)}{1-\np(\nl)} \, .
\end{align*}
Since $0 < \np'(\nl) < +\infty$ for all $0 < \nl \leq 1/2$ and since
\begin{align*}
\lim_{\nl \rightarrow 0^+} \ri (1-\np(\nl)) (\log_2 (1-\nl) - \log_2 \nl) = + \infty \, ,
\end{align*}
there exists $a > 0$ such that
\begin{align*}
\ri(1-\np(\nl)) (\log_2 (1-\nl) -\log_2 \nl) > c\, \np'(\nl), \\
\quad \quad \textrm{for all } 0 < \nl < a \, .
\end{align*}
This latter inequality implies
\begin{align*}
\ri(1-\np(\nl)) (\log_2 (1-\nl) -\log_2 \nl) > c\, \np'(\nl) - \nd \frac{c\, \np'(\nl)}{\np(\nl)}, \\
\qquad \textrm{for all } 0 < \nl < a
\end{align*}
uniformly with respect to $\nd \in [0,1/2)$, which is equivalent to $\partial\, \f(\nd,\nl) /\partial \nl > 0 $ for all $0 < \nl < a$, independently of $\nd \in [0,1/2)$. Therefore the maximum cannot be taken between $0$ and $a$, with $a$ independent of $\nd \in [0,1/2)$.

%%%%%%%%%%%%%%%%%%%%%%%%%%%%%%%%%%%%%%%%%%%%%%%%%%%%%%%%%%%%%%%%
\subsection{Proof of Necessity} \label{sec:necessity}
So far we have proved that the condition on $(\ri,\ro)$ expressed by Theorem \ref{theorem_inner} is sufficient to have a positive normalized typical minimum distance. Now we need to show that this condition is also necessary.  We need to prove that for the ensemble $\msr{C}_{\infty}(\oensemble,\Omega, \ri, \ro)$ all rate pairs $(\ri,\ro)$ such that $\lim_{\nd \rightarrow 0^+}\G(\nd)=0$,  the derivative of the growth rate at $0$ is positive, $\lim_{\nd \rightarrow 0^+} G'(\nd) > 0$.

According to Lemma \ref{lemma:growth_rate_derivative} the expression of $G'(\nd)$ corresponds to
\[
G'(\nd) = \log_2 \frac{1-\nd}{\nd} + \log_2 \frac{\np(\nlo)}{1-\np(\nlo)}  \, .
\]
Hence, since $G'(\nd)$ is the sum of two terms the first of which diverges to $+\infty$ as $\nd \rightarrow 0^+$, a necessary condition for the derivative to be negative is that the second term diverges to $-\infty$, i.e., $\lim_{\nd \rightarrow 0^+}\np(\nlo)=0$. This case is analyzed in the following lemma.
\begin{lemma}\label{lemma:limit_p}
If $\np (\nl)=0$ then $\nl \in \{ 0, 1 \}$ in case the LT distribution $\Omega$ is such that $\Omega_j=0$ for all odd $j$, and $\nl=0$ for any other LT distribution $\Omega$.
\end{lemma}
\begin{IEEEproof}
Let us recall that $\np(\nl)$ is the probability that the LT enconder picks an odd number of nonzero intermediate bits (with replacement) given that the intermediate codeword has Hamming weight $\nl h$. If $\Omega_j > 0$ for at least one odd $j$, then the only case in which a zero LT encoded bit is generated with probability $1$ is the one in which the intermediate word is the all-zero sequence. If $\Omega_j=0$ for all odd $j$, there is also another case in which a nonzero bit is output by the LT encoder with probability $1$, i.e., the case in which the intermediate word is the all-one word.
\end{IEEEproof}

\medskip
Consider now a pair $(\ri,\ro)$ such that $\lim_{\nd \rightarrow 0^+}\G(\nd)=0$. For a fixed-rate Raptor code ensemble corresponding to this pair, we have a positive typical minimum distance if and only if $\lim_{\nd \rightarrow 0^+} G'(\nd)<0$.
%Since from Lemma~\ref{lemma:growth_rate_derivative} we know that $G'(\nd)$ is the sum of two terms, the first of which diverges to $+\infty$ as $\nd \rightarrow 0^+$, a necessary condition for the typical minimum distance to be positive for the considered $(\ri,\ro)$ pair is that the second term diverges to $-\infty$, i.e., $\lim_{\nd \rightarrow 0^+}\np(\nlo)=0$.
By Lemma~\ref{lemma:limit_p} this implies $\lim_{\nd \rightarrow 0^+} \nlo(\nd)=0$ when $\Omega_j>0$ for at least one odd $j$. It implies either $\lim_{\nd \rightarrow 0^+} \nlo(\nd)=0$ or $\lim_{\nd \rightarrow 0^+} \nlo(\nd)=1$ otherwise. That $\nlo(\nd)$ cannot converge to $0$ follows from the proof of sufficiency (as shown, the maximum for $\delta \in [0,1/2)$ is taken for $\lambda > a >0$). To complete the proof we now show that, in the case where $\Omega_j=0$ for all odd $j$, assuming $\lim_{\nd \rightarrow 0^+} \nlo(\nd)=1$ leads to a contradiction.

In case $\Omega_j=0$ for all odd $j$, a Taylor series for $\np(\nl)$ around $\nl=1$ is $\np(\nl) = \bar \Omega (1-\nl) + o(\nl)$. Assuming $\lim_{\nd \rightarrow 0^+} \nlo(\nd)=1$, we consider the left-hand side of \eqref{eq:critical_point} and calculate its limit as $\nd \rightarrow 0^+$. We obtain
\begin{align*}
& \lim_{\nd\rightarrow 0^+} \frac{\partial \f}{\partial \nl} (\nd, \nlo) \\
&= \ri \lim_{\nlo\rightarrow 1^-} \log_2 \frac{1-\nlo}{\nlo} \\
&+ \lim_{\nd \rightarrow 0^+} \! \left( \frac{\nd}{\log 2} \, \frac{\np'(\nlo)}{\np(\nlo)} - \frac{1-\nd}{\log 2} \, \frac{\np'(\nlo)}{1-\np(\nlo)} \right) \\
&= \ri \lim_{\nlo\rightarrow 1^-} \log_2 \frac{1-\nlo}{\nlo} + \frac{1}{\log 2} \lim_{\nd \rightarrow 0^+} \frac{\np'(\nlo)(\nd-\np(\nlo))}{\np(\nlo)(1-\np(\nlo))} \\
&= \ri \lim_{\nlo\rightarrow 1^-} \log_2 \frac{1-\nlo}{\nlo} + \frac{1}{\log 2} \lim_{\nd \rightarrow 0^+} \frac{\bar \Omega (1- \nlo) - \nd}{1-\nlo}
\end{align*}
where the last equality follows from the above-stated Taylor series. According to \eqref{eq:critical_point}, the last expression must be equal to zero, a constraint which requires the second limit to diverge to $+\infty$ (as the first limit diverges to $-\infty$). This, however, cannot be fulfilled in any case when $\nd$ converge to zero and $\nlo$ to one. In fact, using standard Landau notation, when $1-\nlo = \Theta(\nd)$ or $\nd = \mathrm o(1-\nlo)$ the second limit converges, while when $1-\nlo = \mathrm o(\nd)$ it diverges to $-\infty$.

\section{Proof of Theorem \ref{pro:outer}} \label{sec:proof_outer}
The proof consists of deriving a lower bound for $\G(\nd)$ and evaluating it for $\nd \to 0^+$. To derive a lower bound for $\G(\nd)$ we
first derive a lower bound for $\we_{\nd}$. Observing \eqref{eq:we_serial} we see how $\we_{\nd}$ is obtained as a summation over all possible intermediate Hamming weights. A lower bound to $\we_{\nd}$  can be obtained by limiting the summation to the term $\nls=1-\ro$ yielding to
\begin{align}
\we_{\nd n} &\geq  \frac{\weo_{\nls h} \wei_{\nls h,\nd n}}{ \binom {h} {\nls h}} %\nonumber\\
 = \weo_{\nls h} \Q_{\nd n,\nls h}\label{eq:appC:truncation}
\end{align}
where we have introduced
\[
\Q_{\nd n,\nl h}:=\frac{\wei_{\nl h,\nd n}}{ \binom {h} {\nl h}}
\]
representing the probability that the inner encoder outputs a codeword with Hamming weight $\nd n$ given that the encoder input has weight $\nl h$.

Hence, we can write
\begin{align}
\G(\nd) &\geq \lim_{n \to \infty}  \frac{1}{n} \log_2 \weo_{\nls h} \Q_{\nd n,\nls h} \nonumber \\
& =  \lim_{n \to \infty}  \frac{1}{n} \log_2 \weo_{\nls h} + \lim_{n \to \infty}  \frac{1}{n} \log_2 \Q_{\nd n,\nls h} \\
&= \ri \left(\Hb(\nls) - (1-\ro) \right) + \lim_{n \to \infty}  \frac{1}{n} \log_2 \Q_{\nd n,\nls h}
\label{eq:app_upper_G}
\end{align}

We shall now lower bound $\lim_{\nd \to 0^+} \Q_{\nd n,\nl h}$. We denote by

\[
q_{j,\l}:=\Pr\{X_i=0|\hw(\vecV)=\l,\deg(X_i)=j\}.
\]
Note that $q_{j,\l}=1-\pjl$. We have that

\begin{align}
\lim_{\nd \to 0^+} \Q_{\nd n,\nl h} & = \left(\sum_j \Omega_j q_{j,\nl h}\right)^n %\nonumber \\
 \geq  \left(\sum_j \Omega_j \underline{q}_{j,\nl h}\right)^n
\end{align}
with $\underline{q}_{j,\l}\leq q_{j,\l}$ being the probability that the $j$ intermediate symbols selected to encoder $X_i$ are all zero.
For large $h$, we have
\[
\underline{q}_{j,\l}=\left( 1- \frac{\l}{h}\right)^j.
\]
Denoting by $\underline{q}_{\l}=\sum_j \Omega_j \underline{q}_{j,\l}$, we have by Jensen's inequality
\[
\underline{q}_{\l} \geq \left( 1- \frac{\l}{h}\right)^{\avgd}.
\]
We have thus that
\begin{equation}
\lim_{\nd \to 0^+} \Q_{\nd n,\nl h} \geq  \left( 1- \nl\right)^{n\avgd}. \label{eq:appC:Qbound}
\end{equation}

Replacing \eqref{eq:appC:Qbound} in \eqref{eq:app_upper_G} and recalling that $h=n\ri$ we get
\begin{align}
\G(\nd) &\geq  \ri \left(\Hb(\nls) - (1-\ro) \right) + \lim_{n \to \infty}  \frac{1}{n} \log_2 \left( 1- \nls\right)^{n\avgd} \\
&=  \ri \left(\Hb(\nls) - (1-\ro) \right) + \avgd \log_2 \left( 1- \nls\right) \\
&= \ri \left(\Hb(1-\ro) - (1-\ro) \right) + \avgd \log_2 \ro
\end{align}
If we now impose the $\G(\nd)=0$ we obtain:
\[
\phi(\ro)= \frac{\bar \Omega \log_2 (1/\ro)}{\Hb(1-\ro) -(1-\ro)} \, .
\]
This expression is only valid when the denominator is negative, that is,  for $1>\ro>\ro^*$, being $\ro^*$ the only root of the denominator in $\ro \in (0,1)$, whose approximate numerical value is $\ro^* \approx 0.22709$.

\section*{Acknowledgements}\label{sec:Acknowledgements}
The authors would like to to thank Prof. Massimo Cicognani for the useful discussions about the proof of Theorem 3.

%\input{./sections/bibliography}
%\bibliographystyle{IEEEtran}
%\bibliography{IEEEabrv,Raptor}


\end{document}